\documentclass{ar}
\usepackage{apjfonts}
\usepackage{epsfig}
\usepackage{graphics}
\usepackage{natbib,graphicx}

\jname{Xxxx. Xxx. Xxx. Xxx.}
\jyear{2019}

\newcommand{\lsim}{\raisebox{-0.13cm}{~\shortstack{$<$\\[-0.07cm]$\sim$}}~}
\newcommand{\gsim}{\raisebox{-0.13cm}{~\shortstack{$>$\\[-0.07cm]$\sim$}}~}

\newcommand{\apj}{Ap. J.}
\newcommand{\nat}{Nature}
\newcommand{\pasp}{Pub. Astron. Soc. Pac.}
\newcommand{\procspie}{Proc. of SPIE}
\newcommand{\aap}{A\&A}
\newcommand{\apjl}{Ap. J. Lett}
\newcommand{\mnras}{Monthly Not. Roy. Astr. Soc.}
\newcommand{\aj}{A. J.}
\newcommand{\apjs}{Ap. J. Supp.}
\newcommand{\icarus}{Icarus}

\begin{document}

\markboth{Jontof-Hutter}{The Compositional Diversity of Low-Mass Exoplanets}

\title{The Compositional Diversity of Low-Mass Exoplanets}

\author{Daniel Jontof-Hutter,$^1$ 
\affil{$^1$Department of Physics, University of the Pacific, Stockton, California, United States, 95211; \\ email: djontofhutter@pacific.edu}
}


\begin{abstract}
Low-mass planets have an extraordinarily diverse range of bulk compositions, from primarily rocky worlds to those with deep gaseous atmospheres. As techniques for measuring the masses of exoplanets are advancing the field towards the terrestrial regime, from ultra-short orbital periods to Venus-like distances, we identify the bounds on planet compositions, where sizes and incident fluxes inform bulk planet properties. In some cases, measurement precisions of planet masses and sizes are approaching the theoretical uncertainties in planet models.

An emerging picture explains aspects of the diversity of low-mass planets although some problems remain; do extreme low density low-mass planets challenge models of atmospheric mass loss? Are planet sizes strictly separated by bulk composition? Why do some stellar characterizations differ between observational techniques?

As the \textit{TESS} mission begins, low-mass exoplanets around the nearest stars will soon be discovered and characterized with unprecedented precision, permitting more detailed planetary modeling and atmospheric characterization of low mass exoplanets than ever before.
\end{abstract}

\maketitle

\tableofcontents

\section{INTRODUCTION}
Before the discovery of exoplanets, the assumption that our Solar System may be typical prevailed. Although the solar system planets cover a diverse range in bulk compositions, certain bounds appeared logical: small rocky planets inside the ``snow line" and more massive and volatile rich planets farther from the Sun.  
Volatile-poor bodies either formed with chondritic abundance ratios (Venus, Earth, Mars) or required a unique formation mechanism (e.g. Mercury or the Moon).
Large gaps in masses and radii separated rocky planets like Earth from the ice giants Neptune and Uranus, and ice giant from gas giants. 

The former gap has been filled by the \textit{Kepler} mission's unanticipated discovery of thousands of sub-Neptunes with orbital periods less than one year.\footnote{The mission plan anticipated a yield of $\sim$600 transiting exoplanets up to Earth-like distances from their host stars \citep{Borucki1997}} These constitute a class of planet that is absent from the Solar System but that dominate exoplanet demographics within the detection limits of transit surveys. The latter gap has persisted to some extent, with planets from 30 -- 80 M$_{\oplus}$ fairly uncommon, at least among transiting exoplanets. 

In this review, I will focus on planets that are lower in mass than 30M$_{\oplus}$. There is nothing special about this criterion other than the observed diversity among the planets that I include. The regime is particularly interesting since it includes the transition from rocky planets to those that retain deep atmospheres at orbital periods that have been explored. 

I reserve the term ``characterized exoplanets" to refer to transiting exoplanets with measured masses. The transit depth gives a direct measure of a planet's physical size. Thus, measuring the mass of transiting exoplanets permits an estimate of its bulk density. Repeated transits provide a precise orbital period and therefore a measure of orbital distance. 

In this paper, I outline the compositional diversity of characterized low-mass exoplanets. In \S 2, I review the inferences in compositional diversity implied by the orbital periods and radii of transiting exoplanets and the emerging picture of a bimodal distribution of planetary radii and, perhaps, compositions. In \S 3, I discuss the theoretical uncertainties that distinguish models of planetary interiors for a given mass and composition. In \S 4, I review how precise mass and density determinations are achieved, from improvements in stellar characterization, to radial velocity spectroscopy (RV) and transit timing variations (TTV). These techniques characterize planet masses over a limited range of orbital periods and planet sizes due to detection biases. In \S 5, I explore the diversity of well-characterized exoplanets and the range of incident fluxes in which compositional diversity is limited. I highlight the progress of low-mass characterization towards the terrestrial regime, and consider the special subsamples of ultra-short period planets and extreme low density low-mass exoplanets (so called superpuffs). Finally, I discuss the diversity of planet compositions within the same systems and highlight some individual high profile systems with the most precise characterizations. 

This review is written as the \textit{Transiting Exoplanet Survey Satellite} (\textit{TESS}) begins its search for transiting exoplanets all over the sky. The \textit{TESS} Input Catalog includes over one million stars \citep{Stassun2017}, and the mission may well discover the nearest transiting exoplanets to the Solar System, allowing more precise characterizations of low-mass exoplanets than ever before. 

As \textit{TESS} increases the inventory of known low-mass planets, the \textit{James Webb Space Telescope} (\textit{JWST}) will probe the atmospheres of exoplanets that are smaller than has been possible with the \textit{Hubble Space Telescope}, and over a vastly wider spectral range \citep{Greene2016}. This will enable a significantly greater inventory of molecular lines to be characterized at far better precision than is possible today, informing detailed models of exoplanetary atmospheres and compositions.

\section{PLANETARY SIZES}
The \textit{Kepler} mission discovered well over 4,000 transiting exoplanet candidates (\citealt{rowe14,Morton2016}) enabling statistical studies on the nature of planetary systems and their hosts. The period-size distribution has been analyzed in much detail in the context of measuring planet occurrence rates. Even in the absence of measured planetary masses, planetary radii have revealed much about the compositional diversity of low-mass exoplanets. 

\subsection{A Gap in the Planetary Size Distribution}
The size distribution of transiting planets reveals a gap in planet sizes, around 1.6--1.8 R$_{\oplus}$. This gap has been resolved in sharper focus with more precisely measured stellar radii following ground-based spectroscopy \citep{Fulton2017} and astrometric parallaxes from the \textit{Gaia} mission (\citealt{Berger2018,Fulton2018}).  The bimodal size distribution of exoplanets hints at a separation by composition, with planets smaller than $\sim$1.7 R$_{\oplus}$ likely to be rocky. The gap was first identified at short periods as a dearth of sub-Neptunes, attributed to atmospheric mass loss close to the host star (\citealt{owe13,Mazeh2016}). However, since planetary radii grow rapidly with even a tiny mass fraction in the form of a gaseous envelope, 1.6--1.8 R$_{\oplus}$ remains an unlikely size range for a low mass planet even in the absence of atmospheric mass loss \citep{Owen2017}. Indeed, the gap in the distribution of planetary radii is not only observed at short orbital periods but is also detectable at orbital periods from 20 to 80 days \citep{Hsu2018}. 

\subsection{Planetary Sizes Within Systems}
Within multiplanet systems, planet sizes are strongly correlated \citep{Weiss2018}, suggesting that transiting planets within the same systems are inherently similar worlds. However, there is moderate evidence that the outer planets among transiting pairs are larger than their inner neighbors, even accounting for detection biases (\citealt{Ciardi2013, Weiss2018}), consistent with the expectation that inner planets are more likely to have suffered atmospheric mass loss. 

\section{PLANETARY STRUCTURE MODELS}
Precise measurements of planetary masses and radii has permitted meaningful constraints on bulk compositions, with some limitations. 
For any bulk density lower than that of a mixture of metal and rock, there remains a degeneracy in volatile content. If transiting exoplanets formed in situ, they are expected to lack primordial water since high temperature condensation binds oxygen to silicates. (e.g. \citealt{chi13,Chatterjee2015a}). However, if planets formed further out, where water can condense, and migrated inward, then a substantial water component in low-mass planets is plausible. 

Planetary structure models numerical integrate the conditions of hydrostatic equilibrium and mass conservation, using an equation of state to relate pressure to density and temperature deep within the planet. Uncertainties in the equation of state ultimately limit the precision of these models for an assumed composition, even if a planet's mass and radius are known to great precision.

\subsection{Rocky Planets}
Since the Earth's density profile is probed by seismic data, it benchmarks models of rocky exoplanets. In planetary interiors, phase transitions between crystalline states cause discontinuities in the equation of state. For example, the transition from the liquid outer core to solid inner core causes a jump in density of 5\%--- a source of uncertainty for planets with unknown temperature profiles. However, the solid inner core makes only 3\% of the Earth's mass. Thus, since more massive planets are likely hotter, they may have even less of a solid inner core than the Earth, despite the increase in pressure \citep{Zeng2016}. 

Theoretical models of rocky planets up to 10 $M_{\oplus}$ differ by up to $\sim$2$\%$ in radius depending on the equation of state used in the iron core (\citealt{Smith2018}). Equations of state are usually expressed as empirical fits and extrapolations to experimental data, which introduces theoretical uncertainty into planet models. In particular there is a gap between the experimental regime and the theoretical limit of high pressure compression from 0.2 to 10 TPa where the equation of state is unknown and interpolations are adopted \citep{sea07}.

Similarly, theoretical models of rocky planets up to 10 $M_{\oplus}$ differ by up to $\sim$2$\%$ in radius depending on the equation of state used in the mantle (\citealt{for07,Wagner2011,Zeng2016}. In most cases, thermal pressure is ignored and, in models of predominantly iron cores and silicate mantles, the equation of state is largely independent of temperature. The accuracy of this approximation increases at higher pressures, and hence thermal effects are likely negligible for super-Earths. 

For all but a handful of planets that are likely denser than rock, these theoretical uncertainties are still smaller than observational uncertainties in radius and mass. However, as observational uncertainties improve, some assumptions may be tested. For example, the fraction of a rocky planet's mass in the form of an iron core (the core mass fraction) is often assumed to match the Earth (e.g. \citealt{dres15}). However, increased precision in mass characterization may test this: a core mass fraction of 20\% can be distinguished from a core mass fraction of 30\% with measurement precisions of 2\% in radius or 6\% in mass \citep{Zeng2016}. 

The ratios of non-volatiles in the Earth have long been assumed to match chondritic meteorites (with minor deviations), which in turn are closely consistent with elemental abundances in the Sun (e.g., \citealt{Brown1950, Hurley1957}). There is thus an expectation that rocky planets need not have an Earth-like composition if their hosts have different abundance ratios among Fe, Si, Mg, C and O than does the Sun \citep{Bond2010}. Furthermore, the high bulk density of Mercury indicates that abundance ratios in a rocky planet need not match the star. Thus, there remains a wide variety of plausible compositions for rocky exoplanets. 

Differences in the ratio Fe:Mg which largely determines the core mass fraction, cause the most dramatic difference in bulk density (e.g. \citealt{Valencia2006,Unterborn2016}). However, differences in mantle composition have less of an effect on the size of a planet given its mass. For a given planetary mass, a range of plausible Mg:Si ratios could change the planetary radius by $\sim$2\% \citep{Grasset2009}. This percentage declines for more massive rocky planets, where the equations of state for silicates converge at high pressure \citep{Zeng2016}.
 
While theoretical differences in composition based on stellar abundances may be used as a prior in models of planetary structure (e.g., \citep{Dorn2017}), exoplanetary masses and radii are not yet characterized precisely enough for such differences to be tested. These theoretical differences may be informed by observational constraints in the \textit{JWST} era. A small number of objects have been discovered by the \textit{Kepler} and \textit{K2} missions in which asymmetric dips in light curves with variable depth are attributed the dusty tails of disintegrating rocky planets \citep{Rappaport2014}. These objects offer the tantalizing possibility of probing the compositions of dusty ejecta and hence constraining rocky planet interiors \citep{VanLieshout2014}. 

\subsection{Water Worlds}
There is some theoretical uncertainty in the radius of water-rich planets. Models by \citet{for07} of purely water worlds are larger than those of \citet{sea07} or \citet{Zeng2013} by $\sim$10\% for a given mass\footnote{Mass-Radius diagrams of observed exoplanets including these theoretical differences are explored in \S 5, (see \textbf{Figures~\ref{fig:MR}--\ref{fig:MRsmall}}).}. This is obviously less of a discrepancy for any realistic composition of a water-rich planet without an atmosphere, which must include a substantial rocky component (so-called super-Ganymedes). More recent studies like those of \citet{Grasset2009} and  \citet{Zeng2013} give closer agreement on the size of water worlds of a given mass. Shock compression data on on H$_{2}$O give uncertainties concerning the density of ices at a particular pressure to 1--2\% in the regime of sub-Neptune planets and, therefore, uncertainty on the radius of a water-rich planet to a fraction of 1\% \citep{Knudson2012}. 

However, these studies generally neglect the contribution to pressure in the equation of state from thermal effects, which explains some of the discrepancy between the results of \citet{for07} and those of other studies. \citet{Zeng2013} implicitly account for temperature by adopting the pressure-temperature relation for H$_{2}$O along its melting curve. Their model assumes that, at any pressure, fluid states of water will efficiently cool to the melting point of a solid ice phase, and that, at lower temperatures, cooling slows down. The thermal state likely has a dramatic effect on the phases of ice present in the mantle and, perhaps, whether the planet has a significant magnetic field, but it has little effect on bulk density if thermal pressure is negligible \citep{Zeng2014}. \citet{Thomas2016} tested the effect of thermal pressure on the size of a water-rich planet of a particular mass. Their models found smaller radii than \citet{for07}, closely consistent with the models of \citet{Zeng2013}. However, they also found that planetary radii of water worlds are sensitive to the temperature of the water layers. In their models, planets that are 30\% water by mass vary in size by $\sim$2\% in isothermal models ranging from 300 K to 1000 K, and up to $\sim$8\% in size for an adiabatic increase in temperature in depth, and surface temperatures varying over the same range. Thus, for a given mass and water-rich composition, planetary models require either an assumed thermal profile or a detailed cooling model. 

\subsection{Deep Atmospheres}
With the inclusion of an atmosphere, the bulk density of a planet strongly depends on its temperature profile as well as its composition. Thus, model compositions of planets larger than water worlds for a given mass require detailed modeling of an atmosphere's thermal history. Planets cool and contract over time, and thus, the modeled composition is age dependent. An older planet likely requires a denser, deeper envelope (with a smaller core) to fill out its volume than a younger planet of the same bulk density. 

Low-mass planets whose volumes are dominated by H/He gas need not have a significant fraction of their mass in the form of gas. An envelope mass fraction as low as a few per cent can cause a 3 M$_{\oplus}$ planet to have a radius as large as 3 R$_{\oplus}$ for a wide range of ages and incident fluxes \citep{Lopez2014}. The cooling time of the envelope increases with incident flux, but also depends to some extent on the composition, since a mostly rocky exoplanet has a higher heat capacity than one composed mostly of volatiles. Furthermore, a higher bulk content of non-volatiles may include radioactive that have a share of the heat budget that increases over time. 

The size of the planet is also affected by atmospheric mass loss as ionizing X-ray/extreme ultraviolet (EUV) flux heats the highest levels of the atmosphere. An induced hydrodynamic wind removes gas, including heaver elements. Models of X-ray/EUV flux driven mass-loss adopt an energy limited formulation. (In the limit of low X-ray/EUV flux, the hydrodynamic blow-off conditions no longer hold and mass-loss reverts to the less efficient Jean's escape regime.) The efficiency of mass removal is poorly constrained although models are consistent with all but a few observed low-mass exoplanet densities and incident fluxes (e.g., \citealt{lopez12,jackson12}). 

Inferences of a planet's composition require constraints on age and a model for atmospheric mass loss over the planet's lifetime, although most of the evolution likely occurs in the first few 100 Myr. The X-ray/EUV flux declines by a factor $\sim$25 from 100 Myr to 1 Gyr in age \citep{ribas05}. Furthermore, preceding cooling and contraction, a planet's radius is larger and its escape speed is lower at earlier times.

\subsection{Atmospheric Characterization and Bulk Composition}
Atmospheric characterization via transmission spectroscopy cannot directly constrain bulk compositions. A hydrogen-rich atmosphere is expected to have easily detectable traces of H$_{2}$O (or CH$_{4}$ and NH$_{3}$) \citep{Miller-Ricci2009}, but reveals little information about the bulk water content of the planet. 

Similarly, the non-detection of water or any volatile species does not provide an upper limit on the bulk water content of the planet. The low-mass regime has generally been frustrated by flat spectra in the near infrared, interpreted as high clouds or hazes blocking atmospheric transmission [e.g., GJ1214 b (M$_{p} = 6.5$ M$_{\oplus}$, R$_{p} = 2.7$ R$_{\oplus}$) \citealt{Kreidberg2014})].

Despite these limitations, transmission spectroscopy indirectly provides some constraints on the bulk composition, and informs models of the evolution of a planet's radius. The detection of molecular species provide constraints on the atmospheric scale height and mean molecular weight in the atmosphere, and thus the density of the atmosphere at a particular pressure level \citep{Miller-Ricci2009}. A steam atmosphere suffers less mass loss than a H/He envelope because of the increased mean molecular weight and a lower scale height. The lower scale height further reduces the integrated mass loss since a planet with a steam atmosphere is likely less inflated in the first 100 Myr than one with a small amount of H/He (\citealt{Rogers2011,lopez12}). 

The lowest mass exoplanet with a detection of water is HAT-P-26b (19 M$_{\oplus}$, 6 R$_{\oplus}$). Its relatively low atmospheric metallicity implies a higher bulk water content at depth \citep{Wakeford2017}. Progress towards the terrestrial regime should continue with \textit{JWST} \citep{Morley2017}. The small sample of strong water detections in the exoplanet atmospheres hint that clouds and hazes are less likely to suppress water absorption lines in planets that are hotter in equilibrium temperature than 700 K and with stronger surface gravity log$g > 2.8$  ($\gsim$6 m s$^{-2}$) \citep{Stevenson2016}. However, the small existing dataset also shows a correlation in the strength of the water band absorption feature with H/He mass fraction, suggesting an equally plausible cause for the suppression of a transmission signal through higher mean molecular weight \citep{Crossfield2017}.

Although atmospheric characterizations may inform models of the evolution and composition of a planetary atmosphere, the bulk ratio of ices to H/He gas in low density planets is only indirectly constrained. Nevertheless, in select cases it may be possible break the degeneracy in volatile composition. For example, the sub-Venus sized planets of Kepler-444 are not expected to have retained any primordial hydrogen, and hence the detection of hydrogen mass-loss (via Lyman-alpha absorption) would favor a water-rich composition \citep{Bourrier2017}. In contrast, the detection of helium mass loss from a small planet would confirm the presence of an atmosphere \citep{Spake2018}.

\section{CHARACTERIZING LOW MASS EXOPLANETS}
\begin{figure}[h]
\includegraphics[width=4.5 in]{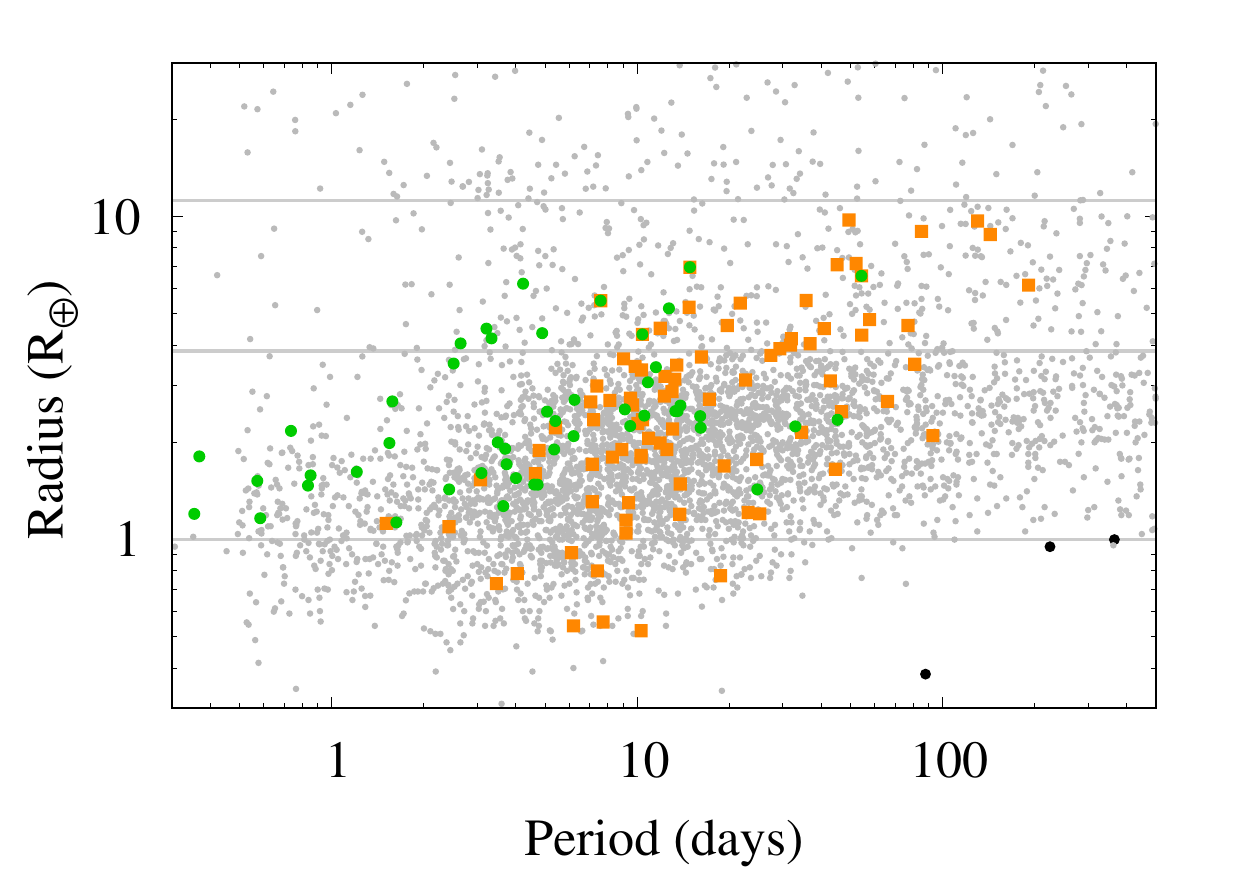}
\caption{Planetary radii and orbital periods of transiting exoplanets in the \textit{Kepler} field (grey points), with the sizes of Earth, Neptune and Jupiter marked with grey lines. Transiting exoplanets with measured masses less than 30 M$_{\oplus}$ are marked with green circles (RV) or orange squares (TTV). Solar System planets are marked as black points. The regimes favored by RV spectroscopy and TTVs complement each other, extending from rocky planets orbiting in less than a day in RV spectroscopy, to Venus-like distances in TTVs, with a some overlap (updated from \citealt{Jontof-Hutter2016}).
}
\label{fig:PR}
\end{figure}

With a measured radius from transit photometry, a precise estimate of a planet's mass is the key to its characterization, constraining its bulk density, surface gravity, escape speed, and the Hill Radius of its orbit. The majority of known transiting exoplanets will not have their masses measured in the near future. However, the subsample that has been characterized represents a fairly wide range of periods and radii among transiting planets. The mass measurements are the combined and complementary results of radial velocity spectroscopy on suitable targets (RV) and transit timing variations (TTV) for multi-transiting systems with favorable orbital periods. \textbf{Figure~\ref{fig:PR}} shows the nominal planetary radii and orbital periods of \textit{Kepler}'s $\sim$4,000 candidate exoplanets. Most are between Earth and Neptune in size. Marked in the figure are the $\sim$130 low-mass exoplanets characterized via RV spectroscopy or TTVs. Both techniques span roughly two orders of magnitude in orbital periods, although RV spectroscopy dominates the mass detections at short periods, and TTVs dominate at longer periods. A few exoplanets benefit from combined mass constraints from RV spectroscopy and TTVs.  

The results of both of these techniques and their detection biases are explored in detail in this section. Both RV and TTV measure the mass ratios of planets to their hosts. Hence we begin by highlighting the progress and challenges in stellar characterization.
\subsection{Stellar Properties}
Much of the progress in characterizing exoplanets is due to the improved precision in measuring stellar masses and radii. The most precise constraints on stellar radius are among nearby stars with directly measured sizes from interferometry (e.g.,  55 Cancri, see \citealt{von11}). However, this technique is limited to targets within $\sim$10 pc, and a larger sample of low-mass exoplanets with directly measured stellar radii awaits \textit{TESS} discoveries. In the \textit{Kepler} field, the same photometric data series that has yielded so many exoplanet discovered also enables the detection of astroseismic oscillations in bright stars either hotter than $\approx$5400 K on the main sequence or evolved \citep{hub13}, allowing precise characterizations of low-mass exoplanet hosts Kepler-36 \citep{car12}, Kepler-93 \citep{dres15}, and Kepler-10 \citep{bat11}. 

Transit photometry directly constrains the bulk density of stars, which enables precise constraints on stellar parameters via astrodensity profiling (\citealt{sea03,Kipping2010,winn11}). The orbital period $P$, transit duraton $T$, transit depth $\delta$, and ingress or egress duration $\tau$ are all measured directly from the light curve and constrain the ratio of the semi-major axis of a small planet's orbit, $a$ to the size of the star, $R_{\star}$:

\begin{equation}
\frac{R_{\star}}{a} = \frac{\pi}{\delta^{1/4}}\frac{\sqrt{T\tau}}{P}\left(\frac{1+e\sin\omega}{\sqrt{1-e^2}} \right),
\label{eqn:arstar}
\end{equation}
where $e$ is the orbital eccentricity and $\omega$ is the argument of periastron. 

Using Newton's rederivation of Kepler's third law and dividing by a star's volume permits the parameters in Equation~\ref{eqn:arstar} to constrain the bulk density of the star:
\begin{equation}
\rho_{\star} \approx \left(\frac{a}{R_\star}\right)^3 \frac{3 \pi}{ G P^2},
\label{eqn:rhostar}
\end{equation}
where $G$ is the universal gravitational constant. Some transiting exoplanets have eccentricities large enough to be detected in RV, although for the existing dataset, low-mass exoplanet eccentricities smaller than $\sim$0.1 are indistinguishable from circular orbits in RV, and in most cases, circular orbits are assumed. Nevertheless, equations~\ref{eqn:arstar} and~\ref{eqn:rhostar} show that even small errors in eccentricity can cause significant errors in inferred stellar densities $\approx$ 3e$\sin\omega$. 

TTVs (\S 4.3) are sensitive to low eccentricities, and hence offer a promising method of reducing this error. Among low-mass exoplanets, there are few detections of significant orbital eccentricity in individual systems. Statistical studies have shown that eccentricities have a narrow distribution in multitransiting systems (\citealt{had14,fab14}). Furthermore, for some individual systems, TTV modeling has constrained orbital eccentricities to $<$0.05. Using TTV  constraints as priors in stellar modeling reduces the uncertainty on the stellar radius (e.g., down to $\sim$2\% for Kepler-11, see \citealt{liss13}).

\citet{Bedell2017} took high resolution spectra to infer improved stellar parameters for the six-planet system Kepler-11. The spectra yielded a stellar density inconsistent at the $\sim2\sigma$ level with that found by fitting the light curve with a photodynamical model (see \S 4.3). The light-curve analysis yielded a consistent measure of $\rho_{\star}$ for all six transiting planets, $\sim$20\% less dense than that found with spectral analysis. This discrepancy is not yet understood, but if densities measured from light curves are systematically underestimated by a similar amount, then some low-mass exoplanets with densities consistent with water may turn out to be rocky. In contrast, if the spectral analysis overestimates $\rho_{\star}$, some low-mass exoplanets with a required envelope may turn out to be have densities consistent with water worlds. However, the majority of characterized low-mass planets (including those at Kepler-11) have low enough densities that substantial volumes of gas are required.

At the time of writing, the stellar parameters of many exoplanet hosts may have improved from the original studies that characterized the exoplanets with additional spectral data \citep{Johnson2017} or with astrometry \citep{Gaia2018}. \textit{Gaia} results measure stars to be $\sim$1.8\% larger than found by astroseismology \citep{Fulton2018}. If this result extends to the hosts of characterized low-mass planets, then it may indicate higher abundances of volatiles in close-in low-mass exoplanets. 

\subsection{Doppler Spectroscopy}
The gravity of an exoplanet causes its host star to revolve around their common center of mass. The radial velocity component from the observer induces a Doppler shift in the spectrum of the host star with semi-amplitude:
\begin{equation}
K = \frac{0.640 \rm{m s}^{-1}}{\sqrt{1-e^{2}}} \frac{m_{p} \sin I}{M_{\oplus}}\left( \frac{m_{\star}+m_{p}}{M_{\odot}} \right)^{-2/3} \left(\frac{P}{1 \rm{day}} \right)^{-1/3},
\label{eqn:RV}
\end{equation}
where $e$ is the orbital eccentricity, $I$ is the inclination of the orbit relative to the sky-plane, $m_{p}$ is the mass of the planet, $m_{\star}$ is the mass of the host star, and $P$ is the orbital period (adapted from \citealt{win11}). For transiting exoplanets, orbital period and phase are measured precisely, and $ I\approx \frac{\pi}{2}$ and thus $\sin I \approx 1$. This resolves the degeneracy in $m_{p} \sin I$ permitting RV spectroscopy characterization of transiting planets. 

\begin{figure}[h]
\includegraphics[width=4 in]{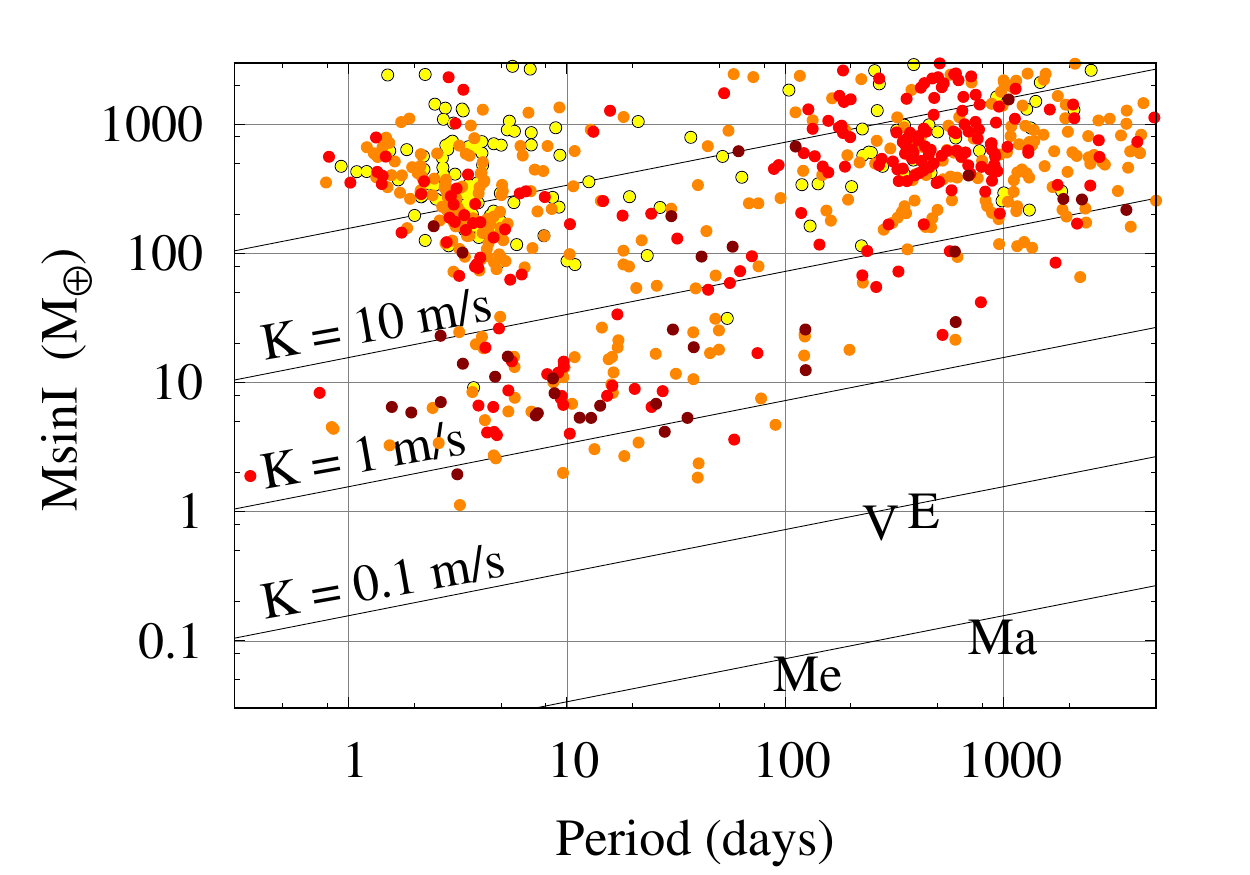}
\caption{Characterized radial velocity (RV) exoplanets, with Mercury (Me), Venus (V), Earth (E) and Mars (Ma) labeled for comparison. Points are color-coded by the spectral type of their host: F (\textit{yellow}), G (\textit{orange}), K (\textit{red}) and M (\textit{maroon}). We include nontransiting planets that have been detected via RV spectroscopy and transiting planets with  RV spectroscopy mass determinations. Curves mark the semi-amplitude (K) for a solar-mass star assuming circular orbits. While the sample is dominated by jovian planets, RV precision is advancing towards the terrestrial regime. These data were generated from exoplanets.org \citep{Wright2011}, with spectral types taken from The NASA Exoplanet Archive (https://exoplanetarchive.ipac.caltech.edu/), the extrasolar planet encyclopedia (exoplanets.eu, \citealt{Schneider2011}) or the Open Exoplanet Catalogue (http://www.openexoplanetcatalogue.com/) }
\label{fig:RV}
\end{figure}

\textbf{Figure~\ref{fig:RV}} shows the orbital periods and measured masses of planets characterized with RV. This includes planets discovered via RV and transiting planets that have been confirmed with RV spectroscopy follow-up. The distribution of planets shown in Figure~\ref{fig:RV} is largely a reflection of detection biases of both RV spectroscopy and transit surveys. Many of the hot Jupiters in the figure were discovered with photometry and confirmed with RV spectroscopy. Close-in planets have a higher transit likelihood---hence the decline in the number of planets confirmed via RV spectroscopy from 10--100 days, compared to 1-10 days. However, the increased density of points near the upper right, beginning around 100 days is due to cool Jupiters occurring much more frequently than hot Jupiters. 

Among low-mass exoplanets, with RV signals below $\sim$10 m s$^{-1}$, there are fewer RV characterizations. The targeted systems are mostly in the Kepler field, where many of the stars are too faint, too hot, or rotate too rapidly for precision RV to find lower SNR exoplanetary signals. Nevertheless, \textbf{Figure~\ref{fig:RV}} clearly illustrates that RV precisions as fine as 1 m s$^{-1}$ are advancing RV spectroscopy to the low-mass regime. Despite the weak dependence on orbital period in Equation~\ref{eqn:RV}, the range of periods where RV signals from low-mass planets are detectable extends to a few weeks, while transiting planets have been detected up to $\sim$1 year in orbital period. No transiting planets less massive than Earth have been characterized using RV spectroscopy, and analogs to the rocky planets of the solar system remain undetectable. For example, an Earth-like planet at 1 AU causes a radial velocity semi-amplitude below 0.1 m/s in a sun-like star, below the detection limit of ground-based observatories (\citealt{Mahadevan2014,Kotani2014}). The lowest mass transiting exoplanets that have been detected in RV spectroscopy include GJ1132 b (1.7 M$_{\oplus}$ orbiting at 1.6 days), Kepler-78 b (1.7 M$_{\oplus}$ orbiting every 9 hours), and GJ9827 b (also known as K2-135 c, 1.5 M$_{\oplus}$, orbiting every 4 days).   

\textbf{Figure~\ref{fig:RV}} also compares RV characterizations by spectral type. While G, K and M stars have planetary mass detections below $\sim$15 M$_{\oplus}$, there are few low-mass exoplanets characterized around F dwarfs. Lower-mass stars have a greater RV amplitude for a given planetary mass and orbital period, which biases mass characterization towards K and M stars (Anglada-Escud\'{e}~et al. 2016). However, for early M-type stars, the noise floor of RV due to stellar activity is typically $\sim5$ m s$^{-1}$ \citep{Gomes2012}. This can be a source of spurious signals and is particularly frustrating to the detection of low-mass exoplanets, since stellar activity and rotation may combine to mimic low signal-to-noise-ratio (SNR) exoplanets with orbital periods of days to weeks (\citealt{Robertson2015, Rajpaul2016, Vanderburg2016}). Nevertheless, the known orbital period and phase of transiting planets permits the planetary signal to be isolated, even if stellar activity causes a greater RV effect than does the exoplanet. This enabled the first characterizations of rocky exoplanets like CoRoT-7 b \citep{leger09}, 55 Cancri e \citep{win11} and Kepler-10 b \citet{bat11}.

At the time of writing, $\sim$50 transiting exoplanets lower  $<$30 M$_{\oplus}$ have been characterized with RV spectroscopy, roughly half in the \textit{Kepler} field (e.g. \citealt{weis13,mar14,dres15}), and an additional 10 have been characterized in the campaigns of the repurposed \textit{Kepler} mission, \textit{K2} (e.g. \citealt{Sinukoff2016,Dai2017,Osborn2017}). 

The sample of low-mass planets characterized with RV spectroscopy will increase substantially with the discovery of small transiting planets around bright stars by the \textit{TESS} mission. Determining the masses of at least 50 additional planets smaller than 4R$_{\oplus}$ in size is one of the primary science requirements of \textit{TESS} \citep{Stassun2017}. The \textit{TESS} catalog prioritizes cooler stars than the \textit{Kepler} Input Catalog, enabling the characterization of low-mass planets over a wider range of incident flux than RV follow-up in the \textit{Kepler} field.

\subsection{Transit Timing Variations}
In multiplanet systems, gravitational perturbations between planets may be detected with transit photometry. A planet may accelerate or slow down from encounters with a planetary neighbor causing transits to deviate from regular periodicity. TTVs are proportional to the mass ratio of the perturbing planet to the host star $\left( m_{p}'/M_{\star}\right)$. In cases where both interacting planets transit, the mass ratios of both planets to the host can be measured from the TTVs, and since their physical sizes are known from the transit depths, both planets' bulk densities are measured.

The first TTV detections were in the form of polynomial trends found in the time series data of measured mid-transit times (\citealt{ford12, stef13}). Dynamical fits (i.e. using N-body simulations) to measured transit times enabled precise mass measurements of sub-Jovians (e.g., Kepler-9, \citealt{hol10}) and sub-Neptunes (e.g., Kepler-11, \citealt{liss11a}) soon after. More directly than fitting measured transit times, photodynamical modeling generates light curves directly with dynamical simulations, fitting for planetary masses, orbital parameters and transit parameters simultaneously \citep{car12}. At the time of writing, dynamical fits to either transit times measured from transit light curves or to light curves directly have characterized $\sim$80 exoplanets $< 30 $M$_{\oplus}$.

Analytical solutions to TTVs has proven invaluable to characterizing exoplanets and understanding the limitations of TTVs. Early analytical progress on TTVs anticipated the detections of non-transiting low-mass exoplanets farther out than observed transiting planets (\citealt{hm05,assc05}) and the detection of exoplanet Trojans \citep{Ford2007}. While these have not yet been realized, several non-transiting Jovian-mass exoplanets orbiting beyond transiting planets have been detected via TTVs and confirmed with RV spectroscopy (\citealt{nes12,Nesvorny2013}). 

\subsubsection{Near-resonant transit timing variations}
\begin{figure}[h!]
\includegraphics[height=3.2 in]{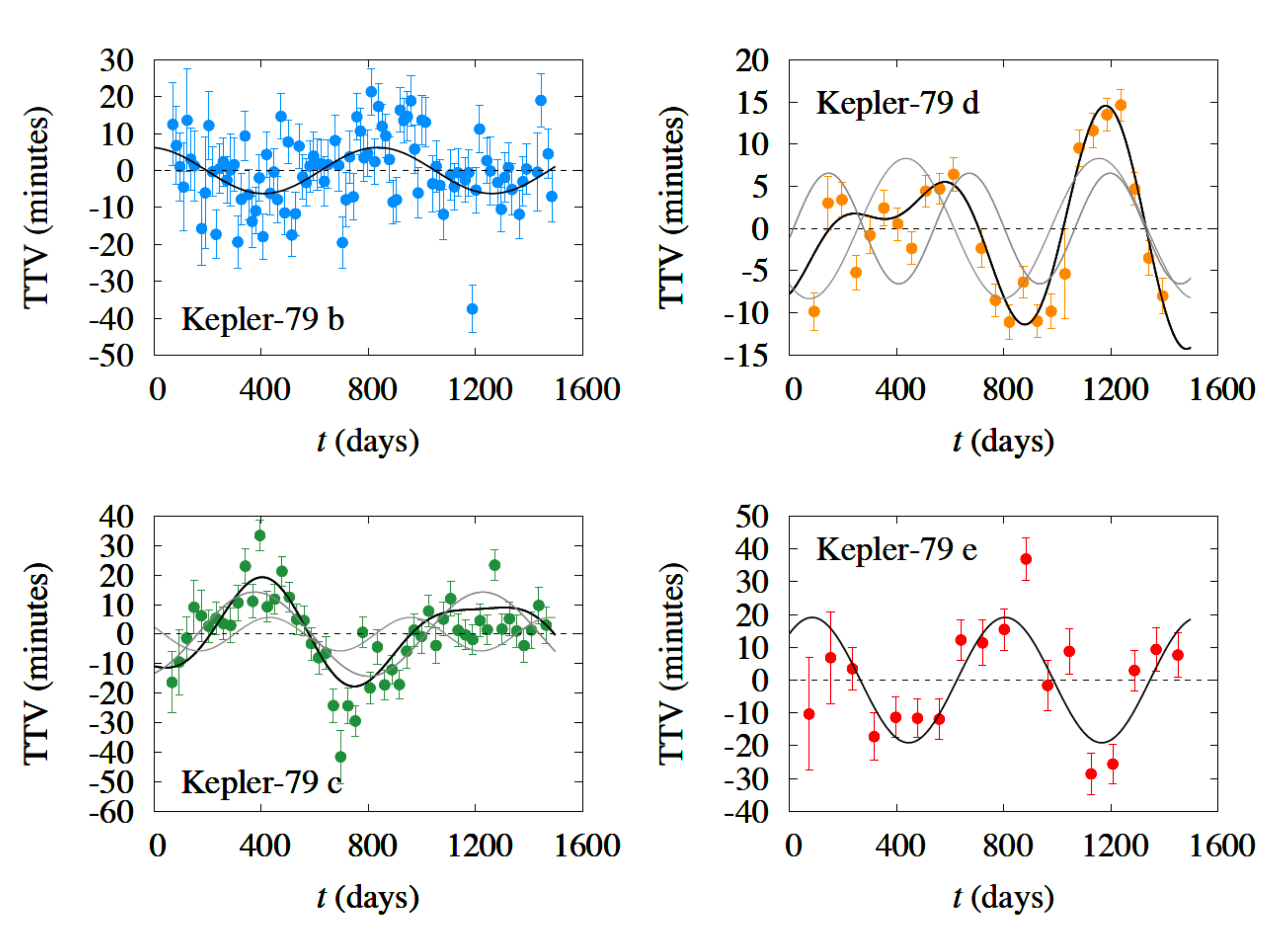}
\caption{Transit timing variations (TTVs)  of Kepler-79 b, Kepler-79 c, Kepler-79 d and Kepler-79 e. The TTVs are the difference between the observed transit times and a linear fit to the transit times that would result from a constant orbital period. The solored points with error bars are measured TTVs. The colored curves are the anticorrelated sinusoidal fits following the solution of \citet{lith12}, with the superposition of two sinusoids (gray curves) for the middle planets, since they are perturbed by two neighbors. Adapted from \citet{jont14}.}
\label{fig:Kep79}
\end{figure}

Most TTV studies have focused on the surprising abundance of systems in which several transiting planets are assembled close to their host star, with orbital period ratios near (but not in) mean motion resonance (such that the ratio of orbital periods is near $j:j-1$, where $j$ is an integer). We illustrate such a system, Kepler-79, which has four known planetary periods near a 1:2:4:6 resonance chain, in \textbf{Figure~\ref{fig:Kep79}}.

The TTV signals of Kepler-79 are dominated by a series pairwise anti-correlated sinusoids, according to the analytical solution of \citet{lith12}. The approximate functional form of this solution are shown in Equations~\ref{Eqn:lith1} and~\ref{Eqn:lith2} (adapted from \citealt{lith12}): 

\begin{equation}
|V| \sim P \frac{\mu'}{\Delta}\left(1+  \frac{|Z_{free}^{*}|}{|\Delta|} \right)
\label{Eqn:lith1}
\end{equation}

\begin{equation}
|V'| \sim P' \frac{\mu}{\Delta}\left(1+  \frac{|Z_{free}^{*}|}{|\Delta|} \right)
\label{Eqn:lith2}
\end{equation}
where $V$ and $V'$ are the amplitudes of the TTV signals of the inner and outer transiting planet, respectively; $P$ and $P'$ are their orbital periods; $\mu$ and $\mu'$ are their masses relative to the host star; $\Delta = \frac{P'}{P} \frac{j-1}{j} -1$ is a measure of how near the period ratio is to resonance; and $Z_{free}^{*}$ is approximately the difference between the free eccentricity vectors (with components $e\sin\omega$ and $e\cos\omega$, where $\omega$ is the argument of pericenter of the planetary orbit) of the interacting planets. 

The periodicity of the TTVs ($P_{TTV}$, known as the superperiod) for near-resonant pairs is predictable from the precisely known orbital periods:
\begin{equation}
\frac{1}{P_{TTV}} = \left| \frac{j}{P'}-\frac{j-1}{P} \right|.
\label{Eqn:Pttv}
\end{equation}

\textbf{Figure~\ref{fig:Kep79}} shows sinusoidal model fits to the TTVs of Kepler-79.  The amplitudes range from $\sim$5 to $\sim$20 min and are readily detectable. The innermost and outermost planets have TTV signals at the expected superperiod given the orbital period ratios between them and their immediate neighbors. The signals of the intermediate planets are the superposition of near-resonant TTVs caused by their two immediate neighbors. 

The TTV amplitudes and phases of transiting planets near resonance, like those shown in \textbf{Figure \ref{fig:Kep79}} yielded many of the early inferred masses of exoplanets more efficiently than did dynamical modeling because of the abundance of multi-transiting systems with period ratios near resonance (\citealt{xie14,had14}). The solution neglects planetary inclinations, with some justification. TTVs are relatively insensitive to mutual inclinations between the transiting planets up to $\sim$50$^{\circ}$ \citep{nes14}. Furthermore, geometry makes significant mutual inclinations between planets in multi-transiting systems very unlikely \citep{Ragozzine2010}. Thus, N-body simulations to TTV data requires orbital periods, phases, eccentricity vector components and planet-star mass ratios. 

However, the eccentricity vector components and mass comprise three unknown parameters per planet in a multi-transiting system and the near-resonant signals provide just two variables, amplitude and phase. Thus, a mass-eccentricity degeneracy exists whereby low-mass planets on eccentric orbits can show TTVs similar to those of more massive planets on circular orbits. In such cases, the ratio of the planetary masses may well be better constrained than the individual masses, since this depends only on the ratio of periods and TTV amplitudes. Furthermore, even where the TTV phases tightly constrain the difference in eccentricity vectors between neighboring planets, individual eccentricities are often poorly constrained, such that adding the same eccentricity vector to an interacting pair of planets has little effect on the TTVs (\citealt{jont15,jont16, Macdonald2016}). In such cases, individual eccentricities are limited only by our prior expectation that, in compact multiplanet systems, they ought to be small.

For planets in resonances, the accuracy of the near-resonant analytical solution fails \citep{Nesvorny2016}. This has little effect on characterized exoplanets, however, since there are few systems with mean motion resonances or resonant chains [e.g., Kepler-223 \citealt{Mills2016}, Kepler-500 \citep{Macdonald2016}, and TRAPPIST-1 \citep{Grimm2018}], and these have all been analyzed with dynamical modeling. 

\subsubsection{Non-resonant transit timing variations}
\begin{figure}[h]
\includegraphics[width=4.5 in]{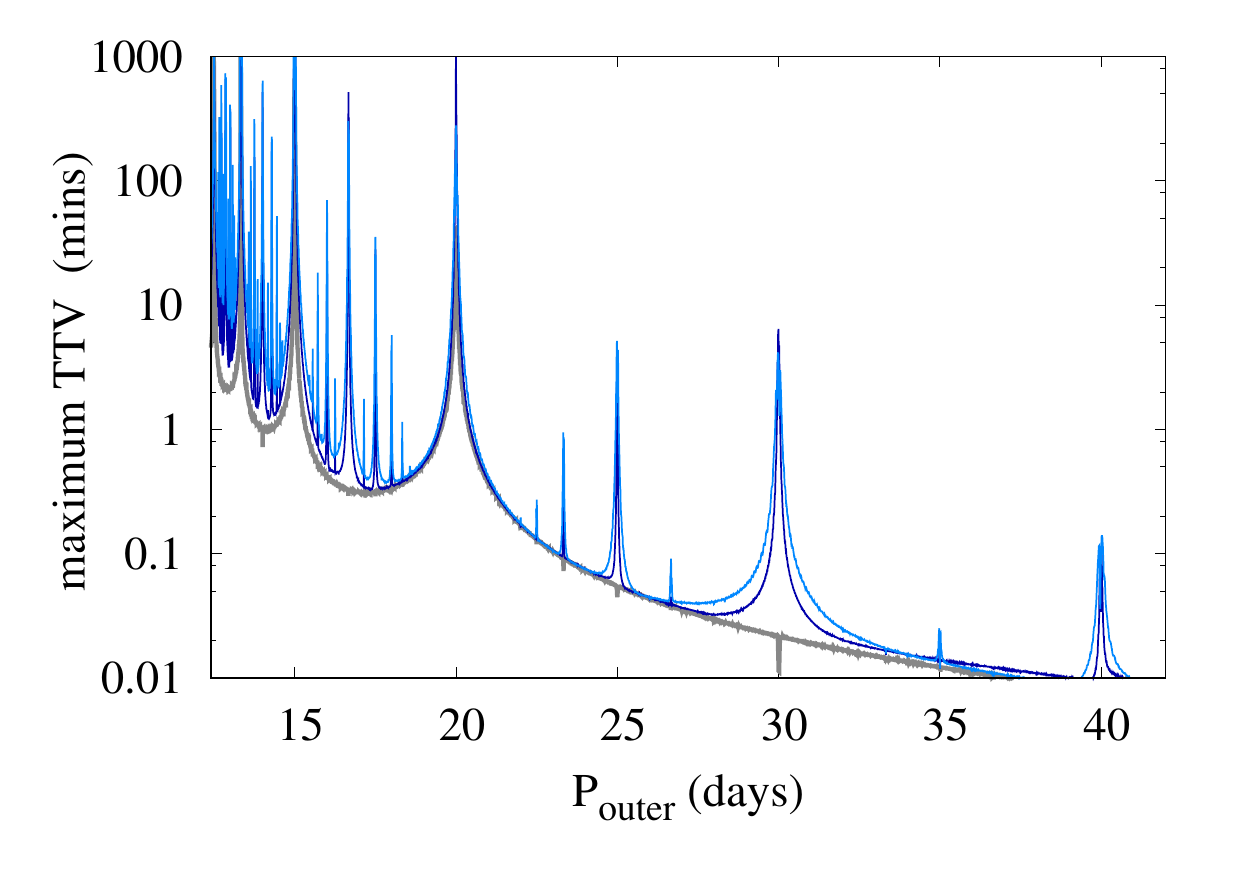}
\caption{Maximum theoretical transit time advancement or delay for an exoplanet on a 10-day circular orbit around a solar-mass star, as a function of the perturber's orbital period. The mass of both planets is 1 M$_{\oplus}$. The thick gray curve corresponds to a perturber with zero initial eccentricity. In blue are models with initial $e\cos\omega$ set to 0.02 (\textit{dark blue}) or 0.04 (\textit{light blue}) with the component $e\sin\omega$ set to zero. Simulated data covered a baseline of 4,000 days.
}
\label{fig:ttvfaster}
\end{figure}

\textbf{Figure~\ref{fig:ttvfaster}} shows the TTVs induced on an inner planet by a perturbing outer planet.\footnote{Since TTVs scale with orbital period and the mass ratio of the perturber to the star, the TTVs of Figure~\ref{fig:ttvfaster} are easily scaled to estimate the TTVs on the inner planet of any planet pair with low eccentricities.} TTVs peak near mean motion resonances, which is why many characterized low-mass exoplanets have period ratios near commensurabilities like 3:2 or 2:1. Between these near-resonant peaks are TTVs primarily caused by synodic interactions between planets \citep{Agol2016}. These are known as chopping because the synodic interactions appear as high-frequency chopping on an otherwise sinusoidal TTV signal. The chopping amplitude declines rapidly with orbital period ratio but causes readily detectable TTVs for planet pairs with close orbital periods. Kepler-36 b and c, for example, orbit with a period ratio closer than the 6:5 resonance and with a strongly chopping TTV signal. \textbf{Figure~\ref{fig:ttvfaster}} also highlights the sensitivity of TTVs to orbital eccentricity, which in most cases increases the TTVs. 

The detection of TTVs requires the transit timing uncertainty to be smaller than the TTV amplitude \citep{nes08}.  In the case of Kepler-79, dynamical fits detect the sinusoidal signals illustrated in \textbf{Figure~\ref{fig:Kep79}} as well as synodic chopping. Kepler-79 d has transit timing uncertainties $\sim$2 min, which is much more precise than the transit timing of its neighbors. It orbits every 52 days, near the 3:2 resonance with its outer neighbor. Scaling the expected TTVs from \textbf{Figure~\ref{fig:ttvfaster}} to the orbital periods of Kepler-79 d and e, we expect a chopping signal $\sim$1.5 minutes even if the planets were far from resonance and Kepler-79 e has a mass as low as 1M$_{\oplus}$. Kepler-79 e's actual mass is 4.1$\pm$1.2 M$_{\oplus}$, and hence there is clearly a detectable chopping component in addition to the near-resonant sinusoidal term in Kepler-79 d. The detected chopping in this system at Kepler-79 d breaks the mass-eccentricity degeneracy and thus permits tight constraints on planetary masses and eccentricities for all four planets. Hence, even though Kepler-79 d's TTVs probe the mass of its neighbors, its precisely measured transit times improves the precision of its mass measurement. 

Note that in cases of weaker-SNR TTVs, the degeneracies remain. However, population studies have shown eccentricities in multi-transiting systems are $\lsim$0.02 (\citealt{fab14,had14}). Such a prior on eccentricity allows for well-contrained masses (e.g. \citealt{jont16, Macdonald2016}).
\subsection{Detection Biases}
As shown in \textbf{Figure~\ref{fig:PR}} a tiny fraction of transiting exoplanets have masses characterized by RV or TTV. While the planets sampled by the two techniques overlap, the separation in sizes and orbital periods of the two techniques highlights their detection biases. 

For RV, the dependence on mass and period in Equation~\ref{eqn:RV} reduces the RV signal-to-noise ratio (SNR) to
\begin{equation}
\rm{SNR}_{RV} \sim \frac{M_{p}}{\sigma_{RV}P^{1/3}},
\label{snr_rv}
\end{equation}
where $M_{p}$ is the planet mass, $P$ is the orbital period, and $\sigma_{RV}$ is the effect of all sources of noise. This simple relation ignores the multiplicity of planets in the system, as well as the stellar properties, and the time required collect data \citep{Steffen2016}. It is clear that RV is biased towards short orbital periods, although fairly weakly. Nevertheless, this dependence on orbital period leaves most transiting planets that are likely low in mass and orbiting at periods longer than days or weeks with undetectable RV signals. Those that are detectable have higher incident fluxes and therefore higher densities in the regime where low-mass planets suffer atmospheric mass loss. Studies like that of \citet{mar14} attempted to correct for this bias by including nondetections in estimation of a mass-radius relation. However, their RV sample was small (49 candidates), and the number of RV detections within the sample even smaller (16 planets). 

TTV characterizations are heavily biased in favor of longer-period planets as show in Equation~\ref{snr_ttv} \citep{Steffen2016}. 
\begin{equation}
\rm{SNR}_{TTV} \sim \frac{M_{p} R_{p}^{3/2}P^{5/6}}{\sigma_{TTV}}.
\label{snr_ttv}
\end{equation}
If we impose the additional effect if a fixed baseline for observations like the \textit{Kepler} dataset, the dependence on orbital period changes to SNR$_{ttv} \sim P^{1/3}$ \citep{Mills2017b}. In either case, the strength of TTV signals increases with period in time series data. Since low-mass planets at longer orbital periods are more likely to retain deep envelopes, they have a wider range of bulk densities (\S 5). Crucially, for a given planetary mass, TTV studies are biased towards larger, low-density planets, since larger planets have deeper transits and more precisely measured transit times.

The differing biases of these techniques are evident in the range of orbital periods and planetary radii of characterized low mass planets, as shown in \textbf{Figure~\ref{fig:PR}}. While the two techniques sample different parameter space in planetary orbital periods and planet sizes, there are several planets with constraints from both RV spectroscopy and TTV [e.g., Kepler-18 \citep{coch11} and Kepler-89 (\citealt{mas13, weis13})]. \citet{Mills2017b} compared the measured masses and found that in 8 of 9 cases, the two techniques yielded consistent results. 

\section{DIVERSE COMPOSITIONS}
\subsection{Mass-Radius Diagram}
\begin{figure}[h]
\includegraphics[width=4.5 in]{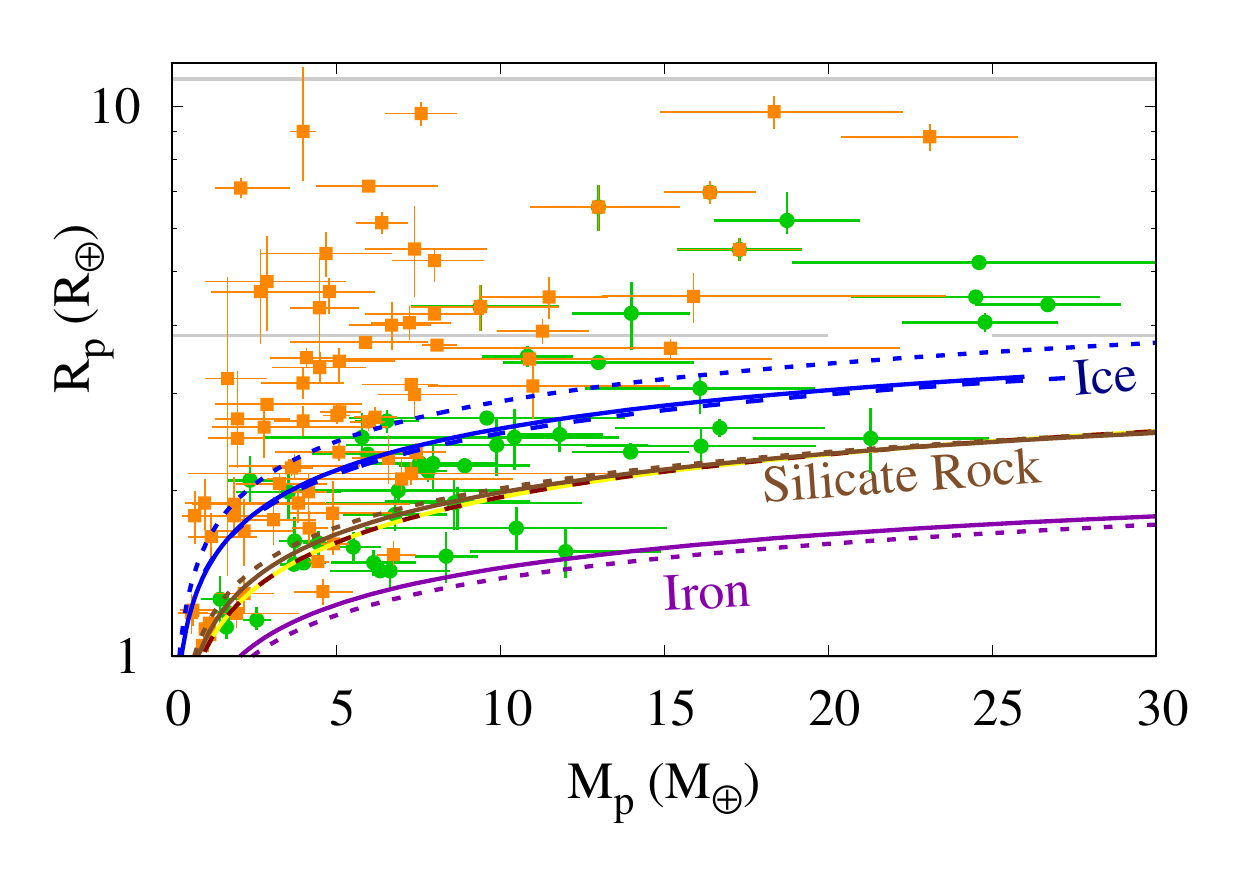}
\caption{Mass-radius diagram for exoplanets below 30M$_{\oplus}$ compared to theoretical models. The solid curves mark mass-radius relations of planets made of pure water ice, rock or iron \citep{Zeng2013}. The short dashes mark solutions from \citet{for07}. The long dashed curves mark models of pure water ($\textit{blue}$) or an Earth-like composition (\textit{maroon} and \textit{yellow}) from \citep{Grasset2009}.Planets characterized from transit timing variations (TTVs) are marked with orange squares; planets characterized from radial velocity spectroscopy (RV) are marked with with green circles. Figure created using data from \citet{mar14} and \citet{Hadden2017} or references therein, with additional data from \citet{Becker2015}, \citet{ofir14}, \citet{Osborn2017}, \citet{pepe13}, \citet{Santerne2018}, \citet{Sinukoff2016}, and \citet{Vanderburg2016}.}
\label{fig:MR}
\end{figure}

A mass-radius diagram of low-mass exoplanets is shown in \textbf{Figure~\ref{fig:MR}} alongside theoretical relations for various compositions. There are some differences in the theoretical models of planets made of pure iron or pure rock based on differences in the equations of state adopted, largely due to the pressure regime from 0.5 to 10 TPa for which there is little experimental data and the theoretical solution in the limit of high pressure is a poor approximation. Various studies adopt different interpolations of the equation of state in this regime although planetary radii are affected by $\lsim$2\% with these differences. For purely water worlds, there is a bigger discrepancy, largely attributable to theoretical uncertainty on the effect of thermal pressure in the equation of state (see \S 3.2). Although a purely water world is an unlikely composition for a planet, the differences between these models are important for planets likely composed of mixtures of rock and water. 

For planets more massive than 10 M$_{\oplus}$, uncertainties in mass are less important than uncertainties in radius for composition modeling. Thus, for planets with precisely measured radii, interior models for a given mass are sensitive to theoretical uncertainties in the density of H$_{2}$O at different depths. At lower masses, measurement uncertainties in mass contribute more to the uncertainty in bulk compositions. 

In addition to characterizing many individual planets, much of the motivation for creating the mass-radius diagram was to determine a relation for using a planet's radius as a proxy for its mass \citep{liss11b}. As more characterized planets have been included on the diagram, various authors have derived mass-radius relations (e.g., \citealt{wu13,Weiss2014}). It is clear from \textbf{Figure~\ref{fig:MR}} that RV spectroscopy and and TTVs are probing different subpopulations by composition. 

Together, RV and TTV mass characterizations leave us with a very wide range in the densities of low-mass exoplanets, with broad scatter in radii for a given mass (or vice versa), rather than a simple one-to-one correspondence (\citealt{Wolfgang2016,Chen2017}). For example, \textbf{Figure~\ref{fig:MR}} shows 54 characterized exoplanets that are nominally between 2 and 7 M$_{\oplus}$ in mass. They range in size from 1.2 to 9.0 R$_{\oplus}$, from super-Earths to Saturns. The low-density regime is overwhelmingly dominated by TTV, while high density planets include some TTV characterization but are predominantly RV. The range of radii for a given mass is attributed to the diversity in compositions for atmospheres that contain a small fraction of a planet's bulk mass, but contributed significantly to its size \citep{Wolfgang2015}.

Planets that are larger than a planet made of pure rock of the same mass must contain volatiles. Furthermore, since it is unlikely that silicates could condense without metals like iron and nickel, it seems reasonable that a planet larger than would be expected from an Earth-like mixture of rock and metal must contain a substantial volatile content. As can be seen in Figure~\ref{fig:MR}, most of the characterized planets  $\sim$5M$_{\oplus}$ are larger than planets made of water, and therefore likely retain deep atmospheres of gas. The actual compositions of these planets are of course, degenerate. Most of the volume must be in the form of gas, but it is unknown how much of the volatile mass could be in the form of water.

The accretion of water during planet formation is not expected inside the snow line, where temperatures are too high and all accreted oxygen is in the form of silicates. Thus, if the observed exoplanets formed in situ, their low densities can be attributed to deep atmospheres of H/He accreted by their rocky cores with little or no water \citealt{Ikoma2012}. \citet{Bodenheimer2014} found that core masses of $\sim$2.5 M$_{\oplus}$ are sufficient for in situ models of gas accretion, but lower masses require formation at more distant, cooler locations (with possible water accretion), followed by migration. The question of where low-mass exoplanets must form is therefore open, as is the related question of whether these exoplanets contain substantial amounts of water. 

The core accretion threshold near 2.5M$_{\oplus}$ is consistent with the characterized low-mass planets: those larger than $\sim$1.6 R$_{\oplus}$ are likely not rocky \citep{rog15}. This apparent separation in compositions at a particular mass and radius, motivated by theory and observed in characterized planets, is also consistent with the hypothesis that the bimodal size distribution is due to composition (see \S 2.1).

\subsection{Composition-Flux Diagram}
As we saw in \S2, there is evidence that the radii of low-mass planets are strongly affected by atmospheric mass loss at short periods. Since luminosity varies by orders of magnitude between stars, insolation is best not measured by orbital period, but rather incident flux or equilibrium temperature. 

The low-mass, low-density planets characterized via TTVs typically have masses below $\sim$6$M_{\oplus}$. However, to be detectable in RV with a semi-amplitude of 2 m s$^{-1}$ around a Sun-like star, such low-mass planets would require an orbital period $<$7 days. For a Sun-like star, this corresponds to an incident flux $\sim$200 F$_{\oplus}$, too hot to retain a deep atmosphere.  
 
Most of the known rocky exoplanets orbit close to their host star, and thus, their dense compositions have motivated atmospheric mass loss models. As methods to compare planet compositions of different masses, planet sizes and bulk densities both have their shortfalls. For masses higher than $\sim$10M$_{\oplus}$, theoretical mass-radius curves flatten rapidly with mass, in which case radius alone can be used as a proxy for composition. However, in the regime $<$5M$_{\oplus}$, planets of similar composition can have very different sizes. 

\begin{figure}[h]
\includegraphics[width=4.5 in]{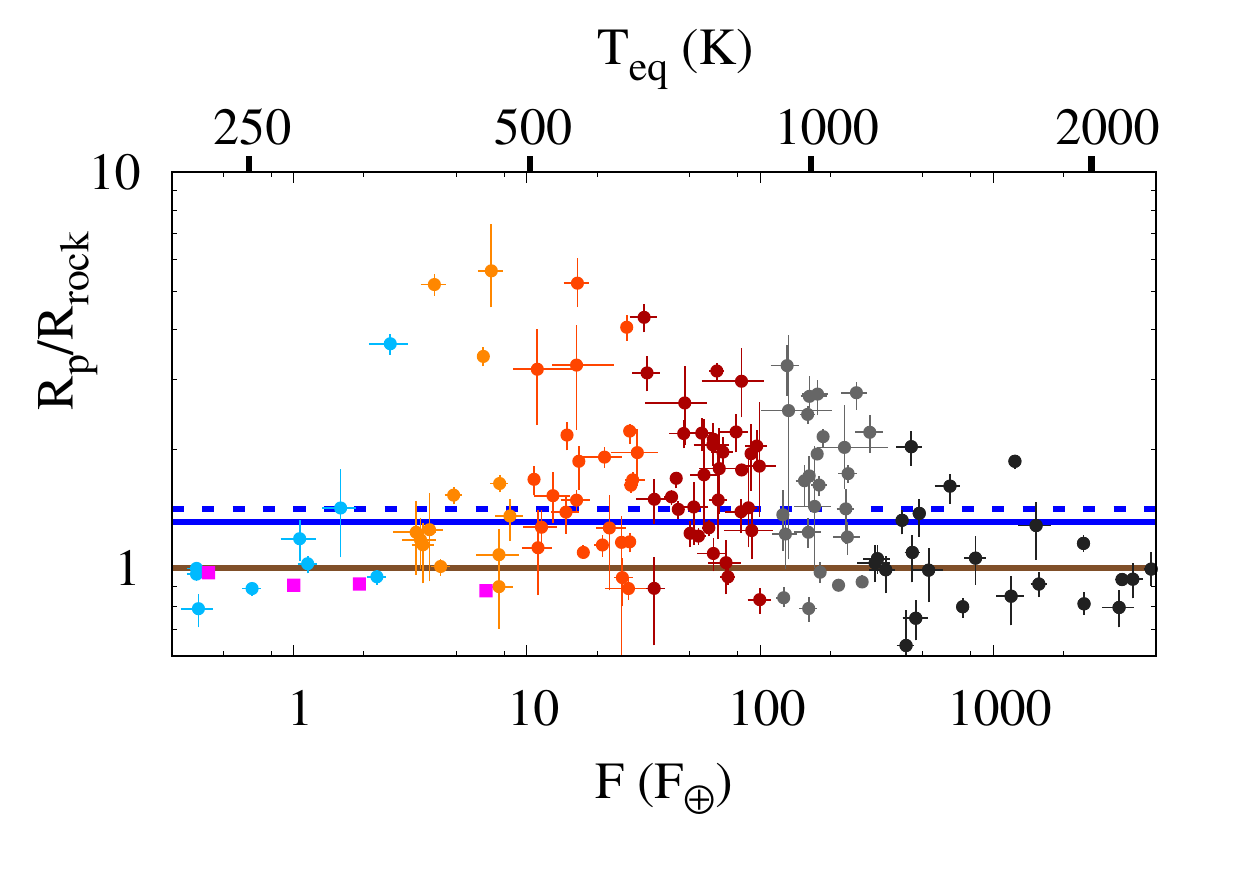}
\caption{Exoplanetary sizes compared to a rocky planet of the same mass (\textit{brown line}) for characterized exoplanets less than 30M$_{\oplus}$. Corresponding blackbody equilibrium temperatures for zero Bond albedo and circular orbits are marked along the top. The solid blue line marks the size of a planet made of pure water ice compared to one of rock, which ranges from 1.32 to 1.30 over the mass range from 1 to 30M$_{\oplus}$ \citep{Zeng2013}. The larger water-world solutions of \citet{for07} follow the dashed blue curve (see \S 3.2). Planets above these line are less dense than water and likely retain deep atmospheres. Colors are coded for incident flux ($F$) compared to Earth's (F$_{\oplus}$): F$<$ 3F$_{\oplus}$ (\textit{cyan}), 3 F$_{\oplus}<$F$<$ 10 F$_{\oplus}$ (\textit{orange}), 10F$_{\oplus}<$F$<$ 30F$_{\oplus}$ (\textit{red}), 30F$_{\oplus}<$F$<$ 100F$_{\oplus}$ (\textit{maroon}), 100F$_{\oplus}<$F$<$ 300F$_{\oplus}$ (\textit{gray}), and F$>$ 300F$_{\oplus}$ (\textit{black}). Adapted from \citet{jont16}.
}
\label{fig:RpRock_flux}
\end{figure}

On the other hand, gravitational compression increases bulk densities substantially with mass compared to their uncompressed values, making density a poor proxy for composition. As a comparison of likely compositions between planets, \textbf{Figure~\ref{fig:RpRock_flux}} scales the sizes of characterized low-mass planets to theoretical models of pure rock, $R_{p}/R_{rock}$, and compares these compositional constraints to their incident fluxes or equilibrium temperatures. The equilibrium blackbody temperature ($T_{eq}$) of an exoplanet, assuming a circular orbit, is
\begin{equation}
T_{eq} = T_{eff}\left(1-A\right)^{1/4}\sqrt{\frac{R_{\star}}{2a}},
\label{Teq}
\end{equation}
where $a$ is the orbital semi-major axis, $R_{\star}$ is the stellar radius, $T_{eff}$ is the blackbody temperature of the star and $A$ is the Bond albedo of the planet. The albedo weaky affects the equilibrium temperature of the planet--a Venus-like Bond albedo as high as 0.77 would only reduce the equilibrium temperature by $\sim$30\%. For low-mass transiting exoplanets, the upper limits of eclipse depths indicate that Bond albedos are$\lsim$0.11 \citep{Sheets2017}. Thus, in \textbf{Figure~\ref{fig:RpRock_flux}}, I adopt the simplest assumption, $A$ = 0.

Planets that are smaller than a purely rocky world of the same mass are denser than pure rock. The exoplanets in this regime all have enough uncertainty in their bulk densities that their compositions are consistent with an Earth-like mixture of metal and rock. \textbf{Figure~\ref{fig:RpRock_flux}} includes a theoretical line that marks the sizes of planets made entirely of water,  since the ratio of the sizes of a planet made of pure water ice to that of one made of pure silicate rock varies by just a few per cent over the mass range 0.3--30 M$_{\oplus}$ \citep{Zeng2013}. Planets below this line in size but larger than planets made of silicate rock must retain a substantial quantity of volatiles, either water or gases, to explain their large volumes. Above this line, planets are less dense than water and must retain deep atmospheres of low density gas, presumably a mixture of hydrogen and helium. 

The largest characterized low-mass planets, with the lowest bulk densities, have incident fluxes between a few and $\sim$100 F$_{\oplus}$.  While cooler low-mass planets appear to have a smaller range in bulk density in \textbf{Figure~\ref{fig:RpRock_flux}}, this end of the graph has a very small sample; characterized low-mass planets in the \textit{Kepler} field are limited to $\lsim$100 days in orbital period, and most of the characterized low-mass exoplanets in the low flux regime are in one system, Trappist-1. 

The decline in planet sizes compared to purely rocky worlds with increasing flux at fluxes greater than $\sim$200F$_{\oplus}$ is consistent with atmospheric mass loss among low-mass exoplanets close to the star. Just how good a proxy incident flux is for integrated high-energy flux that causes mass loss from a planet remains an open question. Nevertheless, it appears reasonable that many of the hot rocky exoplanets that have been characterized are the may be the cores of photoevaporated low-mass planets. However, the sample of characterized exoplanets remains too small to show if a gap in compositions exists at lower incident flux, while the gap in planet sizes extends to longer orbital periods (see \S 2).

\subsection{The Terrestrial Regime}
\begin{figure}[h]
\includegraphics[width=4.5 in]{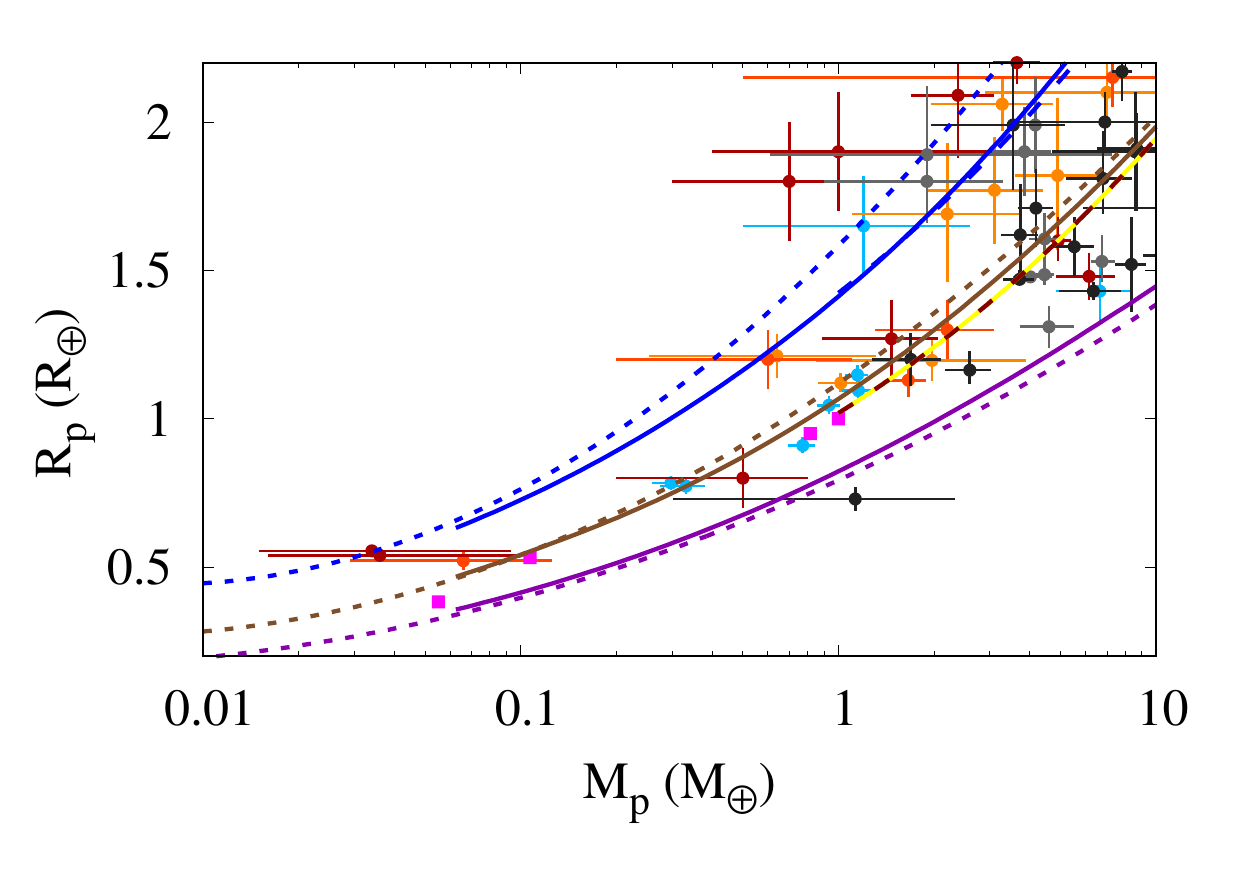}
\caption{Planetary mass-radius diagram of exoplanets below 2 R$_{\oplus}$ compared to theoretical models of planets made of pure water ice, rock or iron (short dashed curves from \citealt{for07}, with solid curves following \citealt{Zeng2013}) and models of pure water or an Earth-like composition (long dashed blue curve or yellow and maroon dashed curve, respectively for $M_{p} > 1$ M$_{\oplus}$ from \citealt{Grasset2009}). Magenta squares mark solar system planets. Colors are coded for incident flux (F) compared to Earth's (F$_{\oplus}$), with F$<$ 3F$_{\oplus}$ (\textit{cyan}), 3 F$_{\oplus}<$F$<$ 10 F$_{\oplus}$ (\textit{orange}), 10F$_{\oplus}<$F$<$ 30F$_{\oplus}$ (\textit{red}), 30F$_{\oplus}<$F$<$ 100F$_{\oplus}$ (\textit{maroon}), 100F$_{\oplus}<$F$<$ 300F$_{\oplus}$ (\textit{grey}), and F$>$ 300F$_{\oplus}$ (\textit{black}).  
}
\label{fig:MRsmall}
\end{figure}

A growing sample of characterized exoplanets smaller than 2 R$_{\oplus}$ allows us to probe the transition from planets with densities consistent with mixtures of rock and metal to those that must retain substantial volumes of gas, as shown in \textbf{Figure~\ref{fig:MRsmall}}. Several exoplanets as small as Mars have been characterized (\citealt{jont15,Mills2017a}). While these planets nominally have densities lower than rock, few of their masses are constrained precisely enough to rule out a primarily rocky composition. 

The most precisely characterized planetary masses up to 6M$_{\oplus}$ have densities well constrained between that of silicate rock and iron, consistent with an Earth-like mixture of rock and iron \citep{dres15}. While several planets shown in \textbf{Figure~\ref{fig:MRsmall}} smaller than 1.5 R$_{\oplus}$ and more massive than Venus appear to be less dense than rock and may be volatile rich, none of these have mass measurements precise enough for a rocky composition to be ruled out. Furthermore, there are very few cases of characterized planets with nominal densities higher than an Earth-like mixture of metal and rock, despite the detection bias towards more massive planets for a given size. Both CoRoT-7 b and Kepler-10 b have a higher iron abundance than Earth for an equation of state that includes thermal pressure \citep{Wagner2012}, while the models of \citet{Zeng2016} give give a core mass fraction similar to Earth for these hot rocky planets. The nearest possible exception, K2-229 b (1.2 R$_{\oplus}$), orbits its host every 0.6 days and appears to have a higher density than can be explained by an Earth-like composition, more closely consistent with a Mercury-like composition of mostly iron \citep{Santerne2018}. 

\subsection{Are there volatile rich planets smaller than 1.5R$_{\oplus}$?}
The pile-up of planet densities near Earth's bulk composition among the most precisely characterized rocky planets is more likely a rough upper limit on bulk densities in the low-mass regime than evidence that the Earth-like composition is the norm. Around 1 M$_{\oplus}$ there is an extraordinary diversity in planet sizes and therefore bulk compositions with some planets even less massive than Earth likely retaining deep atmospheres. While this may seem at odds with the separation in exoplanet sizes (which implies planets smaller than $\sim$1.7R$_{\oplus}$ are rocky), almost all of the characterized exoplanets smaller than 1.5R$_{\oplus}$ have enough uncertainty on their masses that they could be rocky. 

There are a few intriguing cases of small planets with likely volatile compositions. TTV analysis of Trappist-1 reveals planetary densities for TRAPPIST-1 e consistent with an Earth-like composition, while the remaining six planets are nominally less dense than an Earth-like mixture of rock and metal. Nevertheless, these planets have enough uncertainty around their masses and radii to include a possible rocky composition \citep{Grimm2018}. 

Another possible example of a volatile-rich planet smaller than 1.5 R$_{\oplus}$ is in the 3-planet system of the M-dwarf Kepler-138. Two of the planets, Kepler-138  c and d, are slightly larger than Earth and very close in size, although the precise size depends on the stellar radius which is unsettled in the literature (\citealt{kip14,jont15,Almenara2018}). However, the TTVs strongly detect a mass ratio of $\approx$3 between them. This is constrained more tightly than either of the individual masses, and it implies that if the more massive planet, Kepler-138 c is less dense than iron, its neighbor must have substantial water content or a deep atmosphere. \citet{jont15} found that both Kepler-138 c and d are 1.2 R$_{\oplus}$ in size, and hence Kepler-138 d must be on the small side of the bimodal size distribution while likely having a deep atmosphere of H/He gas. \citet{Almenara2018} performed photodynamical model fits to the light curve of Kepler-138 and found different stellar parameters and larger planetary sizes. This potentially rules out Kepler-138 d as an exception to the interpretation that planets smaller than $\sim$1.5$R_{\oplus}$ are likely volatile poor. However, if the innermost sub-Venus or Mars-sized planet Kepler-138 b is larger than measured in earlier studies, it must have a substantial volume of volatiles, either water or gases, to explain its low density. 

\subsection{Ultrashort period planets}
There is some evidence of a trend in \textbf{Figure~\ref{fig:MRsmall}} that the most massive (or densest) planets in the size range 1.5--2R$_{\oplus}$ have higher incident fluxes. This is consistent with the increased diversity of densities at lower incident flux (\textbf{Figure~\ref{fig:RpRock_flux}}). Of  \textit{Kepler}'s 106 planets with orbital periods shorter than one day, all but a few are smaller than $\sim$1.8R$_{\oplus}$ in size \citep{Sanchis2014}. 

Upper limits on the masses of ultra-short period planets, given their proximity to the stellar Roche limit indicate a likely rocky composition \citep{Rappaport2013}. Furthermore, several ultra-short period exoplanets have low masses, as characterized via RV spectroscopy, including  55 Cancri e \citep{Ligi2016}, CoRoT-7 b \citep{Faria2016}, Kepler-10 b \citep{Weiss2016}, Kepler-78 b (\citealt{pepe13,how13}), K2-106 b \citep{Guenther2017}, K2-131 b \citep{Dai2017}, K2-229 b \citep{Santerne2018} and WASP-47 e \citep{Vanderburg2017}; all of these planets orbit G or early-K dwarfs. Both 55 Cancri e and WASP-47 e (both in multiplanet systems with close-in Neptune and Jovian-mass planets) are less dense than would be with an Earth-like composition, hinting at a water-rich composition. 

A significant fraction of ultrashort-period planets are in known multi-transiting systems, and the expectation remains that most of them have undetected low-mass planetary neighbors farther out \citep{Petrovich2018}. Thus, many ultrashort-period planets may be the exposed cores of the missing short period sub-Neptunes \citep{Winn2018}, and their outer neighbors may have similar low densities to those of the majority of characterized sub-Neptunes.

\subsection{Superpuffs}
Most of the compact multitransiting systems discovered by \textit{Kepler} contain low-mass sub-Neptunes, less than 10 M$_{\oplus}$ in mass. Several stand-out low density planets, so called superpuffs, appear in Figure~\ref{fig:MR}, including Kepler-79 d \citep{jont14}, Kepler-87 c \citep{ofir14} and the three lowest density planets known, all orbiting Kepler-51 \citep{mas14}. The size of Kepler-51 c is uncertain because it has a grazing transit. Nevertheless, its transit depth gives a minimum radius, while the TTVs in the system give precise masses. Hence, the upper bound on its density is robust. 

The superpuffs, all $\lsim$0.1 g cm$^{-3}$ in bulk density, are in multiplanet systems and are characterized via transit timing. Their orbital periods range from 45 to 200 days, and thus following Equation~\ref{eqn:RV}, they induce an undetectable radial velocity semi-amplitude $\lsim0.2$ m s$^{-1}$. They range up to Saturn in size, and therefore Saturn-sized objects have a mass range spanning orders of magnitude. The three systems with superpuffs that I highlight in this section are all unique. While all three planets of Kepler-51 are extreme low-density planets, at Kepler-79, three of the four known planets have sizes and masses typical of \textit{Kepler}'s sub-Neptunes, while one is a superpuff. Kepler-87 has two known planets, a superpuff orbiting at 192 days, and a Jovian-mass inner neighbor.

It is important to note here that the observed extreme low densities are not due to the mass-eccentricity degeneracy that occurs in model fits to low-SNR TTV systems. Eccentricities among Kepler's multitransiting systems are known to be small from population studies of transit durations \citep{fab14} and TTVs (\citealt{had14,Hadden2017}). In individual cases like Kepler-79 and Kepler-51, the mass-eccentricity degeneracy is largely broken due to detectable synodic chopping in the TTVs, leaving tightly constrained planet-star mass ratios and eccentricities. 

The extreme low density of these planets is a challenge to models of atmospheric mass loss from low-mass planets (\citealt{lopez12,Owen2016b,Cubillos2017}). In the case of Kepler-51, \citet{mas14} speculates that the likely young age of the system may cause the planets' exceptionally low densities. However, such a solution is unlikely to explain the extreme low density of Kepler-79 d: The system is likely $\sim$3 Gyr in age, and the other three known planets in this system have densities that are typical of compact multi-transiting systems \citep{jont14}. Note that migration models cannot resolve the issue of mass loss for superpuffs since the high energy flux that dominates mass loss models occurs up until a $\sim$100 Myr after formation, while migration models leave planets near their observed orbital distances as soon as the protoplanetary gas disk dissipates within a few Myr. 

Even the most extreme low density low-mass planets are mostly rocky by mass. Kepler-79 d has a mass fraction in its envelope possibly as low as 20--30\% assuming an equilibrium temperature of 500 K, and even lower at higher temperatures (\citealt{Rogers2011}). The formation of these planets is difficult to explain, since such protoplanets should experience runaway gas accretion. However, superpuffs are also undoubtably rare, since only a few have been found with TTVs despite detection biases, and planets larger than Neptune are far less common than planets below this size (see \textbf{Figure~\ref{fig:PR}}). While these planets fairly isolated in the Mass-Radius diagram, it remains uncertain whether they form the readily detectable tail to a continuous spectrum planetary densities or if such planets form in manner quite different to other low-mass exoplanets.

 \citet{Lee2016} offer an in situ formation model for super-Earths in a relatively gas-poor disk. This would explain the abundance of low-mass, low-density planets. However, the most extreme low-density planets are a challenge to in situ formation if the disk is gas poor. In such cases, superpuffs like the three planets of Kepler-51 may be exceptions, accreting their thick atmospheres beyond 1AU and migrating inward. One outstanding difficulty with this is that at Kepler-79, the superpuff is the third of four planets, and the outermost planet has a density typical of \textit{Kepler}'s sub-Neptunes.  
 
The expectation that superpuffs did not form in situ makes them an interesting subsample of low-mass transiting exoplanets. \textit{TESS} may find some extreme low-density low-mass exoplanets at high ecliptic latitudes where there will be a longer photometric baseline \citep{Ricker2014}. However, the occurrence of such planets at longer periods than found by \textit{Kepler}, where they may be cooler and denser, may remain unknown for a long time. They have been characterized exclusively with the TTVs of multitransiting systems of planets near resonances. The non-detectability of these superpuffs in RV makes their existence among single transiting exoplanet systems difficult to determine unless or until \textit{TESS} finds isolated warm sub-Saturn's around bright stars where RV upper limits may indicate extremely low bulk densities. 

\subsection{Diversity within systems}
While planets are typically of similar size within multitransiting systems, there are exceptions, such as WASP-47, where a likely rocky ultrashort-period planet, a hot Jupiter, and a Neptune all orbit within $\sim$9 days \citep{Becker2015}. Among characterized low-mass exoplanets, there are few known differences in composition between planetary companions that cannot be explained by atmospheric mass loss having occurred at the inner, denser planet [e.g., Kepler-10, \citep{Weiss2016} or Kepler-80, \citep{Macdonald2016}]. In the case of Kepler-36, \citet{Lopez2013} invoke substantially different core-masses to explain the different densities between the rocky Kepler-36 b and its extremely close Neptune-sized neighbor Kepler-36 c. In their models, the outer planet was just massive enough to retain its deep atmosphere. Two known low-mass planets have significantly denser low-mass outer neighbors: Kepler-52 b \citep{Hadden2017} and Kepler-105 b \citep{jont16}.  

\subsection{Compositional Diversity and Hosts}
\begin{figure}[h]
\includegraphics[width=4.0 in]{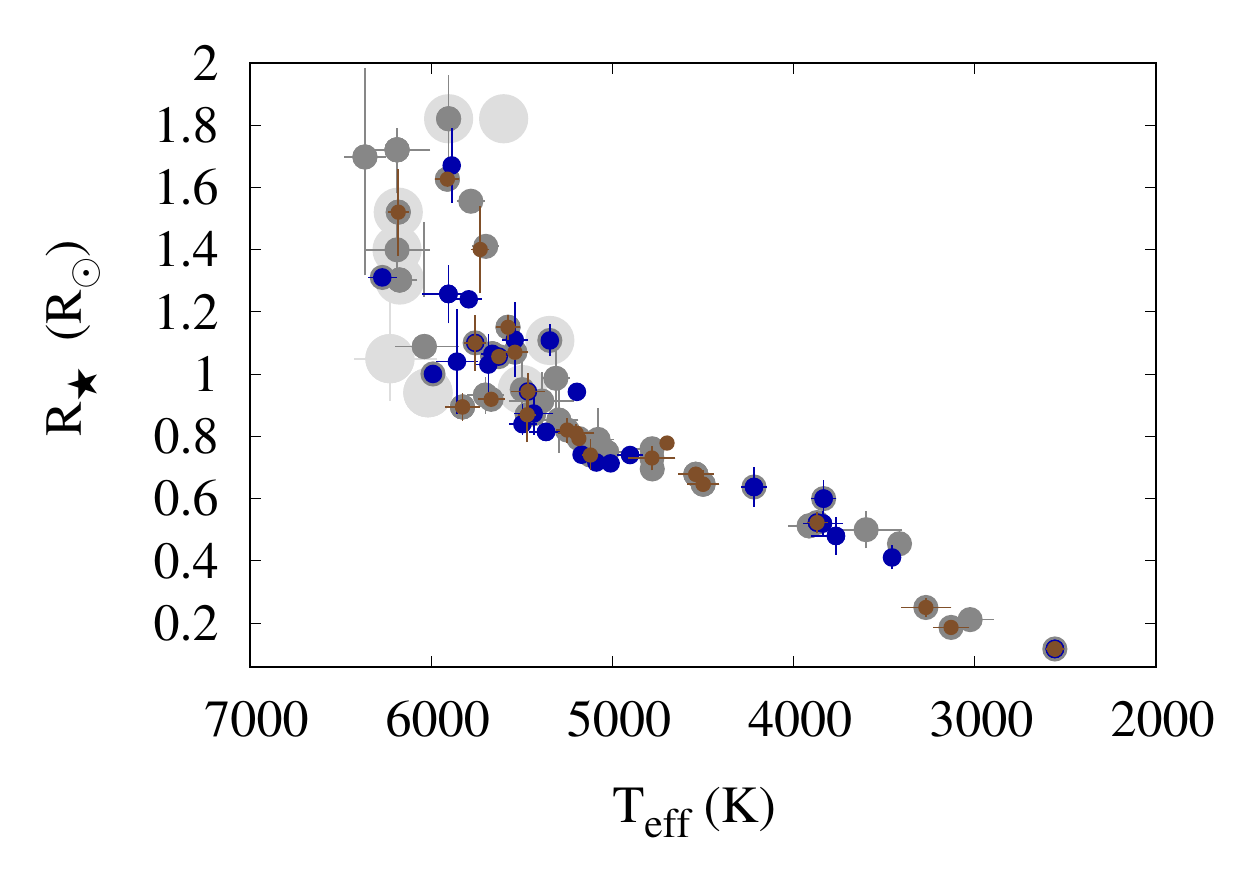}
\caption{The stellar radii and effective temperatures for the hosts of characterized exoplanets $<$30 M$_{\oplus}$, with stellar effective temperature T$_{eff}$ decreasing to the right. Colors are coded by the size of the planet compared to a planet made of silicate rock of the same mass: $R_{p}/R_{rock}$, a proxy for composition. Planets denser than rock  ($R_{p}/R_{rock}$<1) are in brown, between rock and water ($R_{p}/R_{rock}$ from 1--1.32) in blue, $R_{p}/R_{rock}$ from 1.32--3 in dark grey, and of extreme low density with $R_{p}/R_{rock} > 3$ in large light gray points. 
}
\label{fig:TeffRstar}
\end{figure}

Above, I discuss the clear diversity in the compositions of low-mass exoplanets, limited primarily by atmospheric mass loss among low-mass planets close to the star; in this sub-section I highlight the properties of the stars that host characterized low-mass exoplanets. \textbf{Figure~\ref{fig:TeffRstar}} shows the stellar radii and effective temperature of hosts of planets of diverse compositions.\footnote{The stellar properties in Figure~\ref{fig:TeffRstar} are taken from the studies that characterized the exoplanets but that, as mentioned in \S4.1 will soon be revised with results from the \textit{Gaia} mission.}

Characterized low-mass planets orbit stars over a wide range of effective temperatures and sizes. We might expect fewer rocky planets around larger stars due to detection biases since the shallow transits caused by rocky planets are even shallower on large stars. However, it is clear from \textbf{Figure~\ref{fig:TeffRstar}} that rocky planets have been characterized at a wide variety of host stars. There also appears very little difference between the numbers of characterized low-mass planets with densities comparable or less than water as a function of stellar type. However, it appears that the lowest density low-mass exoplanets including all known superpuffs are at hotter or larger stars.

\subsection{Select Systems and Exoplanets}
In this section, I highlight some high profile exoplanets or multi-transiting systems, particularly those in which precise characterizations have enabled insights into the planetary compositions. Extreme low density low-mass planets are explored in more detail in \S 5.6.
 
\subsubsection{Trappist-1}
There are no known solar system analogs to Earth among characterized rocky exoplanets cool enough to have retained a substantial quantity of volatiles like Kepler-138 d but still mostly rocky. However, the transiting planets of Trappist-1 span a similar range of incident fluxes to the terrestrial planets. All seven have precisely measured masses ranging from 0.3 to 1.2 M$_{\oplus}$. The low fractional uncertainties in mass, (all $< 15\%$) are due to the strongly detected TTVs from a resonance chain and detected chopping in six of the planet's TTV signals \citep{Grimm2018}. The most precisely characterized planets, Trappist-1 f and Trappist-1 g are both Earth-sized and have bulk densities less than rock, and likely require $\sim$5\% water by mass or a thick atmosphere, although an Earth-like composition cannot be ruled out. Trappist-1 e is nominally denser than a Earth-like mixture of rock and metal. The resonant configuration of the system is consistent with planet formation further out, followed by migration to their current orbits, all within 0.07 AU of the star.

\subsubsection{55 Cancri e}
55 Cancri has five planets discovered via RV spectroscopy, including two Jovians orbiting at 15 days and 14 years, and two sub-Saturn mass planets orbiting at 44 and 260 days. The star is just 12 pc away, permitting its size to be directly measured with interferometry, in turn permitting a precise measurement of the radius of the transiting planet \citep{von11}. 55 Cancri e transits every 0.74 days and has mass 8.631$\pm 0.495$ M$_{\oplus}$ and size 2.031$^{+0.091}_{-0.088}$ R$_{\oplus}$ \citep{Ligi2016}. Its density is consistent with rock although it appears likely that there are some volatiles. Its short period would cause the rapid loss of a hydrogen atmosphere (T$_{eq} \approx$1900 K), and thus, 55 Cancri e is likely to be water rich \citep{Dorn2017}.

\subsubsection{Kepler-36}
Kepler-36 has two known planets orbiting with a period ratio of just 1.17, causing TTVs of $\sim$5 hours in both planets. The host has detected astroseismology modes enabling tight constraints on the stellar parameters. Kepler-36 b (4.45 M$_{\oplus}$, 1.49 R$_{\oplus}$) remains the most precisely characterized rocky exoplanet, with its fractional mass and radius uncertainties reported at 7\% and 2\% respectively \citep{car12}. Kepler-36 c (8.08 M$_{\oplus}$, 3.68 R$_{\oplus}$) is less dense than Neptune, and must retain a deep atmosphere in addition to any water. The proximity of the two planets to each other and their markedly different densities makes this system a benchmark for atmospheric mass loss models (\citealt{Lopez2013, Owen2016a}). 

\subsubsection{Kepler-4 b}
A hot Neptune, Kepler-4 b (M$_{p}$ = 24.5 M$_{\oplus}$, R$_{p}$ = 4.5$R_{\oplus}$, F$_{p}$ = 1200 F$_{\oplus}$) orbits a slightly evolved star (T$_{eff}$ = 5781 K, R$_{\star}$ = 1.55 R$_{\odot}$) and is large enough that the retention of a deep atmosphere is required. This is unique among planets below 30 M$_{\oplus}$ in mass with incident flux above 700 F$_{\oplus}$. The retention of the atmosphere may be due to its higher mass. Alternatively, since the star has evolved, its current incident flux may be a poor proxy for its integrated EUV/X-ray incident flux.

\clearpage

\section{SUMMARY AND DISCUSSION}
Low-mass exoplanets show extraordinary diversity in bulk composition (\S 5.1). Characterized planets with $\sim$5 M$_{\oplus}$ mass range in size from super-Earths to Saturns. Similarly, at any particular radius from super-Earths to Saturns there is a wide range of plausible masses in the low-mass regime.

In some cases, observational uncertainties in masses and radii are approaching theoretical uncertainties in planetary structure models. At low masses, where radii increase rapidly with mass, the theoretical uncertainty in radius for rocky planets is less important than uncertainties in mass in constraining compositions (\S 3.1). At higher masses compositions are more sensitive to uncertainties in radius. There is some convergence on the theoretical sizes of water worlds in the absence of a gaseous atmosphere, although an assumed temperature profile is necessary (\S 3.2). For larger low-mass exoplanets, compositions are highly degenerate. Even in the absence of water, planetary radii for a given mass and composition requires detailed models of thermal evolution and atmospheric mass loss (\S 3.3). Such models may be informed by additional information from transmission or emission spectroscopy, although the regime of low-mass exoplanet atmospheres awaits the \textit{JWST} era. 

Detection biases in RV spectroscopy and TTVs cause them to be sensitive to complementary regimes in orbital periods, with RV characterizations of low-mass planets primarily at orbital periods shorter than $\sim$1 week, and TTV characterizations over periods of $\sim$1 week to several months (\S 4.4). The RV low-mass planets are more likely to have suffered atmospheric mass loss (\S5.2 and \S 5.4) and include more rocky planets. The TTV characterizations, in constrast, are biased towards planets with longer periods and low densities, including extreme low-density super-puffs (\S 5.5). Some of these provide a test for atmospheric mass loss models and formation models. 

Within systems, there is evidence that neighboring planets have similar compositions, subject to atmospheric mass loss for inner planets. The bimodal size distribution among transiting exoplanets hints of a transition in planetary compositions among planets around 1.5--1.8 R$\oplus$ in size, with planets smaller than this range likely volatile poor. The bimodality must be due to some extent to atmospheric mass loss, but the division extends to longer orbital periods. However, some characterized exoplanets smaller than 1.5 R$\oplus$ may have significant volumes of volatiles, including Kepler-80 e, Trappist-1 f and g, and perhaps Kepler-138 d (or Kepler-138 b, depending on the stellar parameters). It is unknown if the possible volatile content of these exceptions is primarily water or hydrogen gas. 

Resolving this question will initially require more precise stellar parameters and a better understanding of discrepant stellar parameters between techniques. Furthermore, more detailed compositional models including atmospheric characterizations with JWST will further constrain planet compositions. Finally, a larger inventory of systems of characterized planets may allow formation models to test predictions on system architectures and planet compositions from in situ and migration models. 

As the \textit{TESS} mission begins, planet characterization is advancing in precision towards the theoretical precision of planet sizes for a given mass and composition. Improved stellar parameters including ages and metallicities alongside exoplanet atmospheric constraints may motivate more detailed theoretical models of the evolution of planetary radii for low-mass planets with deep atmospheres. Since the input catalog of \textit{TESS} favors M-dwarfs, where small exoplanets are more detectable, there is particular interest in small transiting planets around small, cool stars. The low-mass planets that will pioneer atmospheric characterization with \textit{JWST} will be dominated by M-dwarf systems.  

\begin{summary}[SUMMARY POINTS]
\begin{enumerate}
\item Low mass exoplanets have an extraordinarily diverse range of densities. Planets of a few M$_{\oplus}$ by mass can be super-Earth or sub-Saturn in size.
Planets a few R$_{\oplus}$ by size can have 2--15 M$_{\oplus}$. 
\item The diversity in compositions of low-mass exoplanets declines with incident flux, consistent with atmospheric mass loss among the short period low mass planets leaving dense rocky cores.
\item Models of rocky planets of a given mass and composition lead to a theoretical uncertainty of around 2\% for the radii of rocky planets, and a few per cent for those of water-rich planets. Precise characterizations of low mass exoplanets are approaching these precisions.
\item RV spectroscopy and TTVs characterize planets over a complementary but overlapping range of orbital periods, enabling low-mass planets to be characterized from ultrashort periods $\lsim$1 day to $\sim$200 days, and down to masses as small as Mars'. 
\item In a growing sample, planets in the terrestrial regime are characterizable. The sample will further increase with TESS. 
\end{enumerate}
\end{summary}

\begin{issues}[FUTURE ISSUES]
\begin{enumerate}
\item Discrepancies in stellar properties from spectral data, transit light curves and/or astroseismology must be resolved with the bright exoplanet hosts found by \textit{TESS} and observed with \textit{JWST}. 
\item The extent to which the the size range 1.5--1.8 R$_{\oplus}$ divides planet populations by volatile content may be resolved with a larger inventory of small planets discovered by \textit{TESS} around bright nearby stars. 
\item The extent to which extreme low-density, low-mass planets exist at distances beyond those discovered by \textit{Kepler} or at lower incident fluxes seems unlikely to be determined by \textit{TESS} in its primary mission.
\item \textit{JWST} will observe transmission spectra of low-mass exoplanets in infrared, at higher sensitivity and over a wider spectral range than \textit{HST}. This will likely resolve over which regimes transmission is blocked by clouds and hazes and permit an inventory of atmospheric molecules present in low-mass transiting exoplanets and more detailed models of atmospheres.
\end{enumerate}
\end{issues}

\section*{DISCLOSURE STATEMENT}
The author is not aware of any affiliations, memberships, funding, or financial holdings that might be perceived as affecting the objectivity of this review. 

\section*{ACKNOWLEDGMENTS}
The author thanks Jonathan Fortney, Olivier Grasset, Jakc Lissauer, Mariah MacDonald, Darin Ragozzine, Evan Sinukoff, Scott Thomas, Kevin Zahnle, and Li Zeng for comments that improved this manuscript. This research has made use of the NASA Exoplanet Archive, which is operated by the California Institute of Technology, under contract with the National Aeronautics and Space Administration under the Exoplanet Exploration Program.

%
\noindent

\end{document}